\documentclass[runningheads]{llncs}

\usepackage{graphicx}

\usepackage{amsmath,amssymb,amsfonts}
\usepackage[inline]{enumitem}
\usepackage{algorithm}
\usepackage{graphicx} 
\usepackage{textcomp} 
\usepackage{comment}
\usepackage{wrapfig}
\specialcomment{new}{\begingroup\color{black}}{\endgroup}
\specialcomment{red}{\begingroup\color{red}}{\endgroup}
\specialcomment{rev}{\begingroup\color{green}}{\endgroup}
\usepackage{array}
\newcolumntype{P}[1]{>{\centering\arraybackslash}p{#1}}
\usepackage{listings}
\usepackage{algpseudocode}
\usepackage{csquotes}
\usepackage{eso-pic}
\usepackage{multirow}
\usepackage{float}
\usepackage[caption=false,font=footnotesize]{subfig}
\usepackage{booktabs}
\usepackage{url}
\usepackage{soul}
\usepackage{enumitem}
\newlist{steps}{enumerate}{1}
\setlist[steps, 1]{label = S.\arabic*:}
\usepackage{longtable}
\usepackage{multirow,tabularx}
\newcolumntype{Y}{>{\centering\arraybackslash}X}
\usepackage{balance}



\begin{document}
\title{On the Integration of Blockchain and SDN: \\ Overview, Applications, and Future Perspectives}
\titlerunning{On the Integration of Blockchain and SDN}
%
\author{Anichur Rahman\inst{1} \and
Antonio Montieri\inst{2}\textsuperscript{,}\thanks{Corresponding author.} \and
Dipanjali Kundu\inst{1} \and
\\
Md. Razaul Karim\inst{3} \and
Md. Jahidul Islam\inst{4} \and
Umme Sara\inst{1} \and
\\
Alfredo Nascita\inst{2} \and
Antonio Pescap\'e\inst{2}}
\authorrunning{A. Rahman et al.}
%
\institute{Department of Computer Science and Engineering, National Institute of Textile Engineering and Research (NITER), Constituent Institute of Dhaka University, Savar, Dhaka-1350, Bangladesh\\
\email{\{anis\_cse,dipanjali\_kundu,ummesara\}@niter.edu.bd}
\and
University of Napoli Federico II, Italy\\
\email{\{antonio.montieri,alfredo.nascita,pescape\}@unina.it}
\and
Mawlana Bhashani Science and Technology University, Tangail, Bangladesh\\
\email{razaulce15004@gmail.com}
\and
Green University of Bangladesh, Dhaka, Bangladesh\\
\email{jahidul.jnucse@gmail.com}}
\maketitle  
\begin{abstract}
Blockchain (BC) and Software-Defined Networking (SDN) are leading technologies which have recently found applications in several network-related scenarios and have consequently experienced a growing interest in the research community.
Indeed, current networks connect a massive number of objects over the Internet and in this complex scenario, to ensure security, privacy, confidentiality, and programmability, the utilization of BC and SDN have been successfully proposed.
In this work, we provide a comprehensive survey regarding these two recent research trends and review the related state-of-the-art literature.
We first describe the main features of each technology and discuss their most common and used variants.
Furthermore, we envision the integration of such technologies to jointly take advantage of these latter efficiently.
Indeed, we consider their group-wise utilization---named BC-SDN---based on the need for stronger security and privacy.
Additionally, we cover the application fields of these technologies both individually and combined. 
Finally, we discuss the open issues of reviewed research and describe potential directions for future avenues regarding the integration of BC and SDN.

To summarize, the contribution of the present survey spans from an overview of the literature background on BC and SDN to the discussion of the benefits and limitations of BC-SDN integration in different fields, which also raises open challenges and possible future avenues examined herein.
To the best of our knowledge, compared to existing surveys, this is the first work that analyzes the aforementioned aspects in light of a broad BC-SDN integration, with a specific focus on security and privacy issues in actual utilization scenarios.
\end{abstract}
%
%
%

\section{Introduction}

With the growing amount of always-online connected devices, the challenges to face within modern network environments have also grown.
It is estimated that nowadays $30$ billion devices are connected over the Internet and this number will reach $75$ billion worldwide by $2025$~\cite{mostarda2021fast}.
This increase of devices is creating a massive number of issues such as attacks against the networking systems, data theft, addressing issues, sensors' energy consumption and battery loss, etc.
Moreover, after the revolution of Industry 3.0 and 4.0~\cite{aceto2020industry}, the Internet is no longer only a medium to exchange files or emails but it is the enabler of several safety-critical operations which makes these issues much more urgent.
Thus, the handling of such safety-critical operations must be carried out to properly utilize the data generated by several devices interconnected over the network~\cite{talari2017review}.

Software-Defined Networking is an interesting paradigm that can be used to enhance and manage various security aspects in modern networks such as Internet of Things (IoT) environments.
For instance, to provide better system security, it is necessary to permit the use of resources only among authorized users~\cite{vandana2016security}.
This limits the possibility that third-party users get control over the system and lowers the frequency of attacks~\cite{karmakar2020sdn}.
Further, the large number of connected devices creates a huge amount of data passing from one system segment to another, producing enormous traffic in turn. As conventional networking devices such as routers or switches have to make choices and then monitor the traffic flow, the overall speed of operations is slower. 
In this sense, the SDN paradigm with the OpenFlow protocol allows the SDN controller to link to other devices and separates hardware from software~\cite{kreutz2014software}.
Indeed, the SDN controller can incorporate a constructive or reactive mechanism to remove or even change the traffic flow via a flow table. The transmitted traffic is then handled in the so-called control layer~\cite{al2016software}, which can also protect the networking devices from internal and external attacks.

Blockchain is another technology that can be a possible solution to enforce a verification system at every edge and handle trust-management issues to make the system robust against various attacks~\cite{hu2020blockchain} and to ensure block validation using encryption or consensus mechanisms~\cite{hu2020securing}.
BC provides a Peer-to-peer (P2P) communication system that maintains a database for all members of the network where every record about the connection establishment is easily stored and properly maintained.
In addition, the chain of blocks can not be easily modified: changes are possible when proper validation is ensured, and only in this case a new block can be included in the BC~\cite{mendiboure2018towards}. Furthermore, it ensures authenticity and confidentiality to the data transmitted among network nodes.

Given the high interest that BC and SDN technologies have generated in the research community (and more recently also in the industry), our research aims to provide insights and guidelines to foster their fruitful integration (i.e. BC-SDN).
More specifically, in this survey, we seek answers to the following research questions:
\begin{itemize}
    \item[\textbf{RQ1.}] \textbf{BC-SDN Ecosystem:} What are the key features and benefits of the BC-SDN integration as investigated in state-of-art works?
    \item[\textbf{RQ2.}] \textbf{Security and Privacy in BC-SDN:} Which are the main security and privacy threats affecting BC and SDN? How have they been faced in literature by exploiting the integration of both technologies?
    \item[\textbf{RQ3.}] \textbf{Application of BC-SDN:} Which are the most relevant applications that can benefit from BC-SDN?
    \item[\textbf{RQ4.}] \textbf{Future Challenges:} What are the challenges to deal with in the near future for integrating BC-SDN with emerging technologies in computational intelligence?
\end{itemize}

\subsection{Related Surveys}
\label{subsec:related}

\begin{table*}[t]
\renewcommand{\arraystretch}{1.05}
\caption{Related surveys regarding SDN and BC. The works are grouped based on the related technology and reported in chronological order within each group.}
\label{tab:T1IoT}
\centering
\scriptsize
\begin{tabular}{@{}llll@{}}
\toprule

\textbf{Tech} & \textbf{Ref.} & \textbf{Year} & \textbf{Main Topic} \\
\midrule

\multirow{8}{*}{SDN} & \cite{yurekten2021sdn} & 2021 & Solutions to cyber defence on SDN based systems with taxonomy. \\

& \cite{ahmad2021scalability} & 2021 & Security and Privacy concerns of Controllers of SDN. \\

& \cite{hasneen2022survey} & 2021 & SDN based Security for 5G framework. \\

& \cite{eliyan2021and} & 2021 & Survey of Solution to DDoS and DoS attack in SDN. \\

& \cite{ray2021sdn} & 2021 & SDN-based systematic review for edge and cloud computing in IoT. \\

& \cite{balasubramanian2021sdn} & 2021 & SDN framework for smart industrial IoT environment. \\

& \cite{younus2019survey} & 2019 &
Integration of SDN and Smart Building.\\

& \cite{sahay2019application} & 2019 & Application of SDN for improving security in computer networks. 
\\

& \cite{farris2018survey} & 2018 & 
SDN-NFV framework for ensuring security in IoT environment.\\

& \cite{wang2018novel} & 2018 &
Security aspects and open challenges of SDN technology.
\\
\midrule

\multirow{16}{*}{BC} & \cite{berdik2021survey} & 2021 & Management and Improvement of information systems security using BC.\\

& \cite{bhutta2021survey} & 2021 & Analysis of the evolution, security of the BC technology. \\

& \cite{hewa2021survey} & 2021 & Aspects of BC with future research scopes. \\

& \cite{9548956} & 2021 & A survey of the application of BS specificially in decentralization method.\\

& \cite{latif2021blockchain} & 2021 & BC for addressing challenges and security aspects in IIoT.\\

& \cite{da2021embedding} & 2021 & Integration of IoT and BC for enhancing security. \\

& \cite{ahmed2020chapter}  & 2020 & Improvements in Industry 4.0 by integrating BC technology.\\

& \cite{li2020survey} & 2020 & Approaches for enhancing security in BC technology.\\

& \cite{sengupta2020comprehensive} & 2020 & Security issues of IoT and solutions provided by BC technology.\\

& \cite{dasgupta2019survey} & 2019 &  
Classification of security vulnerabilities in BC technology.\\

& \cite{feng2019survey} & 2019 & Privacy issues associated with BC-based applications.\\

& \cite{gao2018survey} & 2018 & Applications and challenges of BC for IoT and other paradigms.\\

& \cite{nguyen2018survey} & 2018 & Proof- and voting-based algorithms for consensus in BC technology.\\

& \cite{salman2018security} & 2018 & BC-based approaches for different security services.\\

& \cite{ferrag2018blockchain} & 2018 & Overview of BC technology and classification of its threat models.\\

& \cite{banerjee2018blockchain} & 2018 & BC-based security techniques designed for, or applicable to IoT.\\
\midrule

\end{tabular}
\end{table*}

The present survey discusses essential information related to both SDN and BC considering also their security and privacy aspects as well as their applications.
In the last years, many works have focused on these technologies proving that they are heavily attracting the interest of the research community.
Hereinafter, we review the most recent surveys (published within the last five years) regarding such technologies and their integration, briefly discussing the aspects on which each work is focused.%
\footnote{When a work covers different technologies, we categorize it based on its main topic.}

\paragraph{SDN.}
In~\cite{barakabitze20205g}, the authors discuss the 5G network softwarization and slicing strategy based on SDN and Network Function Virtualization (NFV) technologies. Different industrial initiatives and projects, their requirements, and various architectural approaches for 5G networks are also described.
Similarly, Bannour et al.~\cite{bannour2017distributed} focus on the SDN approach and particularly on distributed SDN controllers.
Improving the security of the SDN environment and especially of SDN controllers is the main topic of the work in~\cite{iqbal2019security}.
Additionally, Bizanis et al.~\cite{bizanis2016sdn} propose the joint utilization of SDN and network virtualization technologies to bring several functionalities to IoT applications.
In~\cite{karakus2017survey}, a brief discussion on SDN control plane focusing on scalability issues is presented. The authors also develop different controller design schemes such as flat, hierarchical, and hybrid controllers.
Furthermore, in~\cite{iqbal2021pcss}, the authors present a strategy for smart homes using SDN that ensures the privacy of the system. The authentication of user is accomplished by encryption procedure based on a symmetric key protocol.
In another SDN-IoT integration-based work~\cite{rahman2021sdn}, the authors suggest to integrate SDN and IoT to develop an intelligent framework that is capable of providing solutions to different problems related to Industry 4.0 applications that have arisen during the Covid-19 pandemic.

\paragraph{Blockchain.}
Tariq and al.~\cite{tariq2020blockchain} propose to leverage BC to manage several applications such as Wireless Sensor Networks, Vehicular Ad-hoc Networks (VANets), IoT, and healthcare systems.
They review the existing problems, their solutions, and possible security issues.
Besides, Deepa et al.~\cite{deepa2020survey} integrate these studies presenting different approaches for potential integration of Big Data applications with BC technology.
More recently, the survey in~\cite{yaqoob2021blockchain} discusses different case studies of practical usage of BC for healthcare data management systems.
The integration of BC within IoT systems is analyzed in~\cite{dai2019blockchain} where the authors
propose the definition of Blockchain of Things to name the synthesis of BC into 5G and industrial applications.
Also, the survey in~\cite{berdik2021survey} focuses on the management of information system using BC with particular attention on the security issues.
Similarly, Bhutta et al.~\cite{bhutta2021survey} present the evolution of BC technology along with the specific security measures introduced, while Hewa et al.~\cite{hewa2021survey} discuss the technical aspects of BC focusing on its future scopes.

\paragraph{Integration of BC and SDN.}
Wadhwa et al.~\cite{wadhwa2021survey} investigate
healthcare systems enriched with the integration of BC and SDN. They describe various use cases where the benefits of these two technologies can be fruitfully leveraged.
With a specific eye on security, in~\cite{hassan2021survey}, the authors discuss SDN-based Intrusion Detection Systems in BC applications.
Likewise, Alharbi~\cite{alharbi2020deployment} takes into account BC to protect the SDN environment, also assessing the feasibility of this powerful integration.

\vspace{10pt}
For the sake of completeness, other related surveys on BC and SDN---not detailed in the present subsection for brevity---are reported in Tab.~\ref{tab:T1IoT}, which groups them based on the related technology and summarizes the main topics covered by each work.

\paragraph{Positioning of Our Survey.}
In the present survey, we aim to investigate the integration of BC and SDN technologies with an eye on security and privacy issues in real applications.
To the best of our knowledge, this study is the first survey that takes into account BC and SDN technologies by providing details on both their main features and deepening their effective integration along with actual use cases.
As reported in previous paragraphs and further summarized in Tab.~\ref{tab:T1IoT}, we analyze in a systematic manner different aspects scattered across previous works.
Indeed, unlike the works investigating the application of BC or SDN alone, we firstly provide an overview of BC and SDN applications underlining their security and privacy aspects, then we discuss the BC-SDN integration by analyzing the peculiarities and (security) issues deriving from it.
In this regard, comparing our survey with recent ones covering BC-SDN~\cite{wadhwa2021survey,hassan2021survey,alharbi2020deployment}, ($i$) we expressly make BC-SDN the main pillar of our investigation (differently than~\cite{hassan2021survey}) and we do not narrow our dissertation to ($ii$) specific use cases (e.g., healthcare~\cite{wadhwa2021survey}) or ($iii$) particular aspects of interest (e.g., implementation feasibility~\cite{alharbi2020deployment}).

\subsection{Contributions and Organization of the Survey}
\label{subsec:contributions}

\begin{figure}[t]
    \centering
    \includegraphics[width=0.6\textwidth]{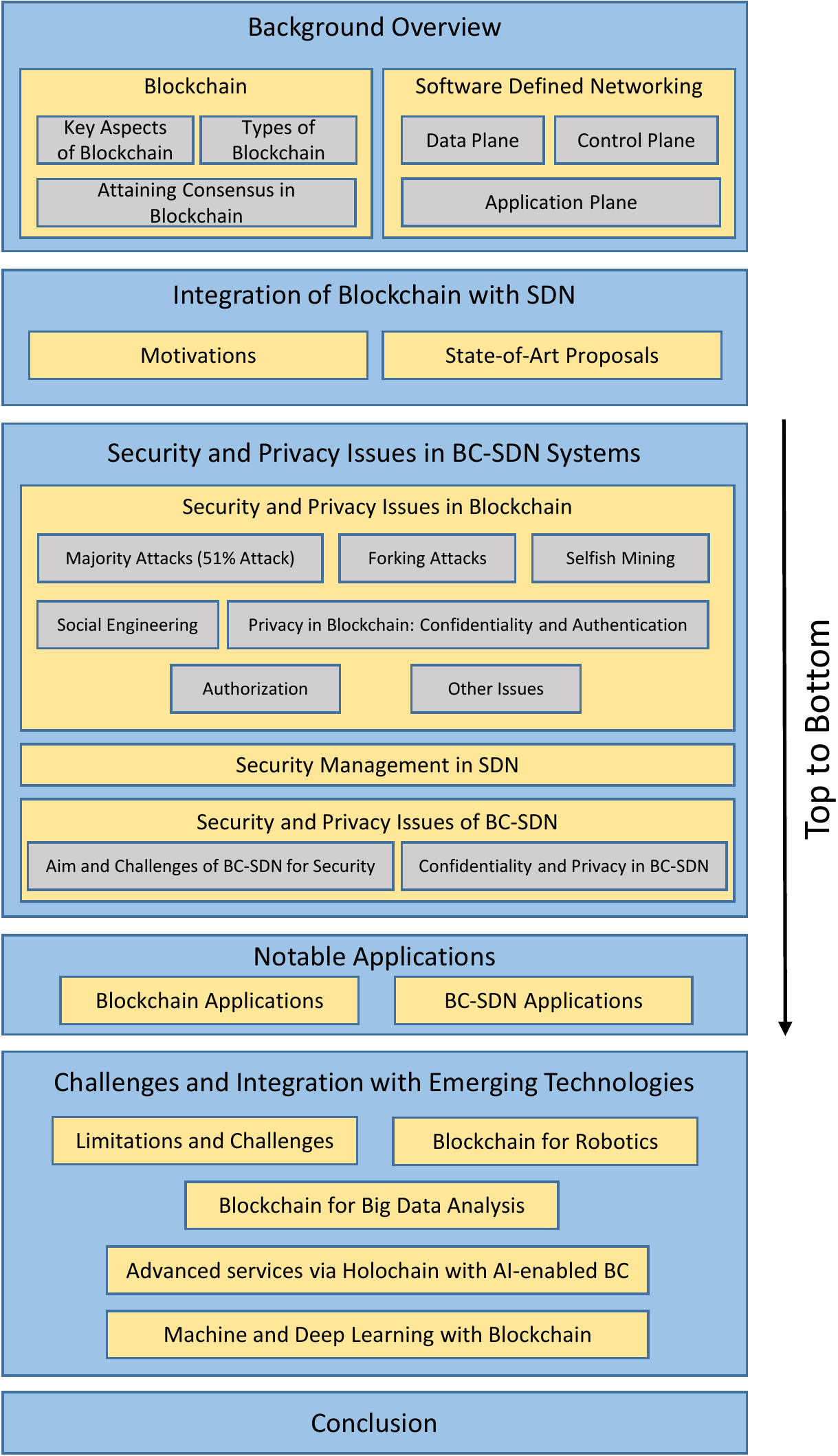}
    \caption{Road map of the present survey.}
    \label{fig:road_map}
\end{figure}

\begin{table}[t]
\renewcommand{\arraystretch}{1.05}
\caption{List of acronyms in alphabetical order.}
\label{tab:acronyms}
\centering
\scriptsize
\begin{tabular}{@{}ll@{}}
\toprule
\textbf{Acronyms} & \textbf{Definitions}\\
\midrule
\emph{ACL} & Access Control List\\
\emph{AI} &  Artificial Intelligence\\
\emph{API} &  Application Programming Interface\\
\emph{BC} & Blockchain\\
\emph{DDoS} & Distributed Denial of Service\\
\emph{DL} & Deep Learning\\
\emph{IDS} & Intrusion Detection System \\
\emph{ML} & Machine Learning\\
\emph{NFV} & Network Function Virtualization\\
\emph{P2P} & Peer to Peer\\
\emph{PoS} & Proof of Stake\\
\emph{PoW} & Proof of Work\\
\emph{SC} & Smart Contract\\
\emph{SDN} & Software-Defined Networking\\
\emph{SPBFT} & Simplified Practical Byzantine Fault Tolerance\\
\emph{VANET} & Vehicular Ad hoc Network\\
\bottomrule
\end{tabular}
\end{table}

In this work, we discuss SDN and BC leading technologies and their fruitful integration. In detail, we provide an extensive analysis of their features, security and privacy issues, and applications. Finally, we discuss open challenges and future perspectives. We also consider valuable benefits achieved with the integration of these technologies in order to better support applications in many fields.
To summarize, in this survey, we offer the following contributions:

\begin{itemize}
    \item We discuss the state-of-art literature and provide a general overview about BC and SDN.
    \item We present the effective benefits of the integration of SDN with BC; we also investigate its existing and upcoming features.
    \item We discuss security and privacy issues of BC and SDN and their integration (i.e. BC-SDN).
    \item We review the applications of these technologies in different fields.
    \item Finally, we describe issues, open challenges, and future investigations related to considered technologies.
\end{itemize}

In view of these aims, we review recent papers related to BC and SDN, and their integration by considering state-of-art proposals (with related motivations) and focusing on their features and benefits, as well as security and privacy concerns.
Moreover, we review articles related to real application scenarios of BC-SDN as an extension of their individual application fields.
In particular, we have selected such studies prioritizing more relevant ones (i.e. those deepening security and privacy aspects or describing actual use cases) published in the last five years (see e.g. Tab.~\ref{tab:BC-App}).

\vspace{10pt}
The remainder of this survey is organized as follows: Section~\ref{sec:overview} presents a general overview of analyzed technologies (i.e. BC and SDN).
Section~\ref{sec:integration_sdn_bc} introduces the motivations behind the BC-SDN integration also reviewing state-of-art proposals.
The security and privacy issues in BC and in the aforementioned integration are discussed in Sec.~\ref{sec:security_privacy},
Successively, the applications of BC and BC-SDN are presented in Sec.~\ref{sec:applications}.
Furthermore, in Sec.~\ref{sec:FD}, we focus on the current research challenges and future directions for the effective employment of considered technologies.
Finally, we conclude the survey in Sec.~\ref{sec:conclusion}.
To ease readers, Fig.~\ref{fig:road_map} depicts the road map of the present survey.
Also, Tab.~\ref{tab:acronyms} summarizes the acronyms used in the text for readability.

\section{Background Overview}
\label{sec:overview}

This section focuses on some key aspects of BC and SDN providing an useful overview of these technologies.
Section~\ref{subsec:blockchain} describes different types of BC, its key aspects, and methods for attaining consensus in transactions.
In Sec.~\ref{subsec:sdn}, the SDN paradigm is discussed, describing the properties of the three planes (i.e. data, control, and application) SDN is made of.

\subsection{Blockchain}
\label{subsec:blockchain}

\begin{figure}[t]
    \centering
    \includegraphics[width=0.64\textwidth]{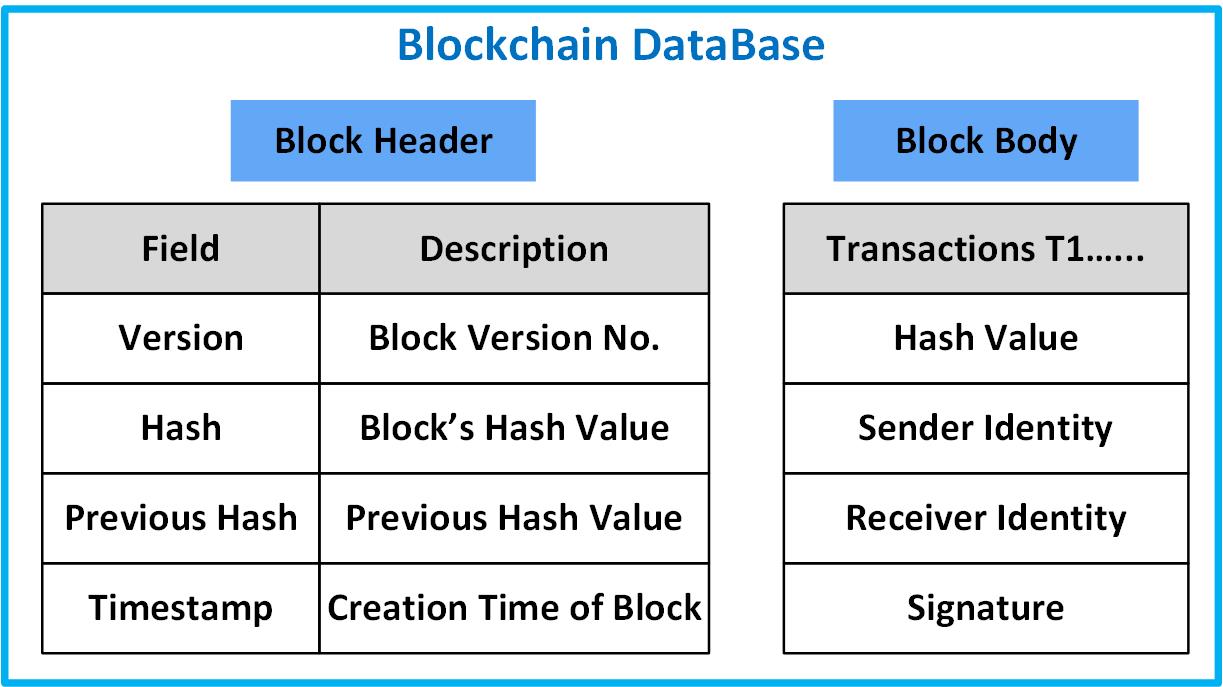}
    \caption{Structure of each block constituting a Blockchain.}
    \label{fig:bct}
\end{figure}

\begin{figure}[t]
    \centering
    \includegraphics[width=0.78\textwidth]{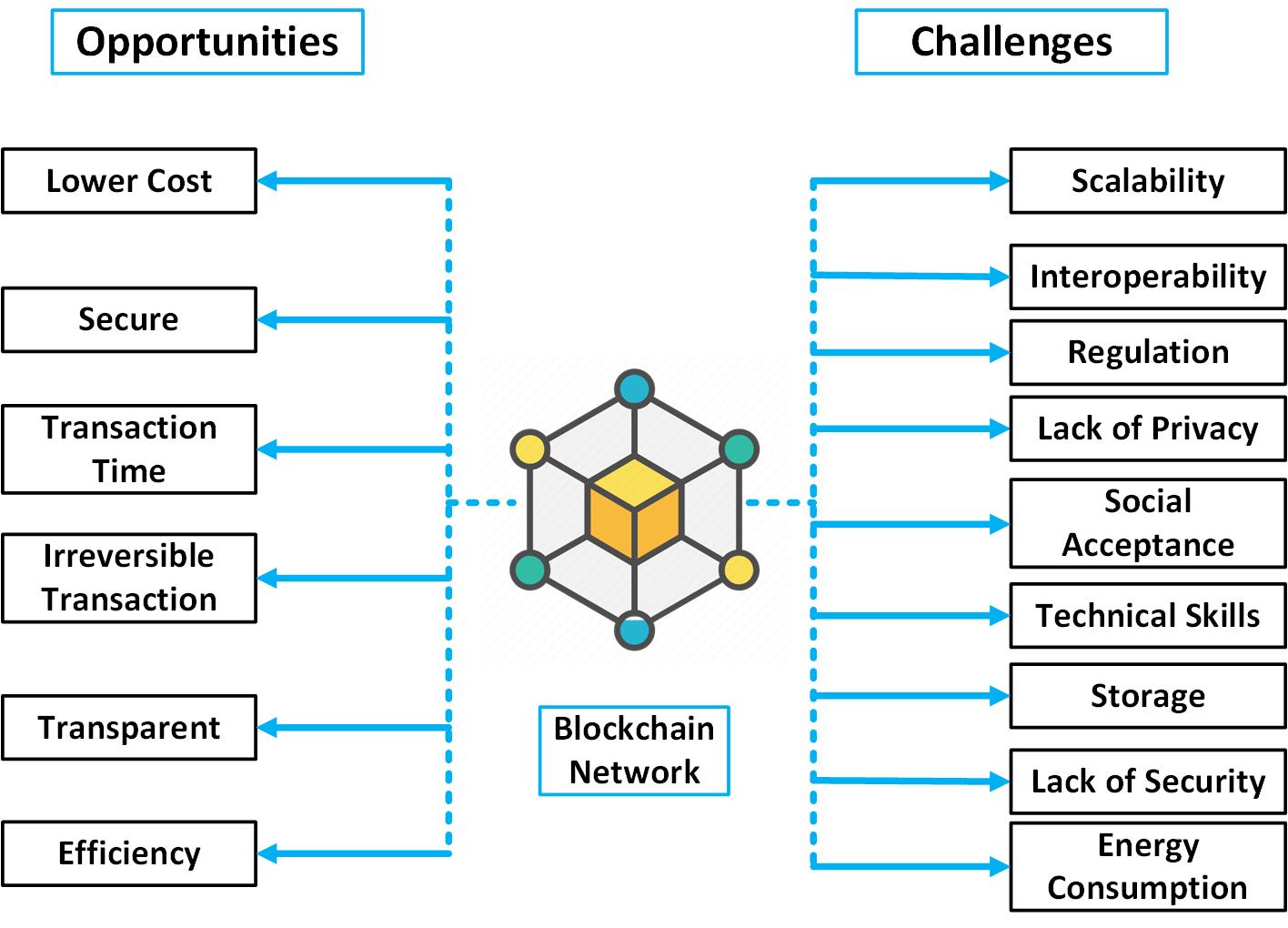}
    \caption{Opportunities and challenges in Blockchain network.}
    \label{fig:bcsdnt}
\end{figure}

BC technology was firstly introduced by Satoshi Nakamoto (pseudonym) in $2008$ and it is currently exploited in numerous applications.
A BC is a decentralized and distributed ledger and every participant can authenticate it without the interference of any central authority or special individual which provides a clearinghouse service verifying and clearing all transactions.
Therefore, the BC is assembled independently by every node in the network.
Each \emph{block} records some or all most recent transactions that have not been recorded in a previous block. As shown in Fig.~\ref{fig:bct}, each block consists of
(i) the block header encompassing current and prior cryptographic block hash values, a timestamp, and a nonce;
(ii) the main body encompassing transaction hash value, sender and receiver identity, and signature for each transaction.
The users who share their computing power to verify if the transactions in the blocks are legitimate in exchange for a reward are called \emph{miners}.
They have to resolve a statistical problem based on the calculation of a cryptographic hash function to confirm the latest transactions and list all of them within the global ledger.
Normally, a block is extracted every $5$ to $10$ minutes.

The newly mined block is attached to the BC once the majority of its participants agree that it is valid; in other words, adding a new block requires to reach a \emph{consensus} among the miners (also called nodes in this context).
More specifically, the consensus is an artifact of the asynchronous interaction of thousands of independent nodes, all following simple rules.
The most common consensus algorithms are the \emph{Proof of Work (PoW)} and the \emph{Proof of Stake (PoS)} in which the miners needs to provide a certain proof that must be validated by other nodes in the network to be allowed to publish (cf. Sec.~\ref{subsec:consensus}).

It is worth underlining that blocks can only be added and not modified.
Indeed, after a block is added to the BC, modifying data in a block of the chain is an extremely challenging task because it requires changing all of the following blocks. Therefore, the BC ledger becomes more and more immutable as time passes.

To summarize, in Fig.~\ref{fig:bcsdnt}, we provide a diagram reporting the opportunities (shown on the left side) and the related challenges (shown on the right side) related to the utilization of the BC technology.
We can notice that if on the one hand, the opportunities and advantages of utilizing the BC are enormous, on the other hand obstacles and challenges must be carefully considered and mitigated to fully exploit BC potential especially in dynamic and controllable environments (as for instance in software-defined networks).

\subsubsection{Key Aspects of Blockchain.}
Hereinafter, we describe the main features that characterize a generic BC.

\paragraph*{Decentralization.}
BC has a distributed and decentralized structure~\cite{rahman2020distb} where there is no central node or trust authority to store data or determine the transaction validity and order. Moreover, BC does not apply any set of rules to establish the transactions or regulating the way nodes interact. Otherwise, consensus is reached via the interplay of thousands of independent nodes independently verifying a set of criteria.

\paragraph*{Transparency.}
BC is transparent in recording new data and also in updating them because the system itself validates and authenticates transactions. Third party or malicious users can not include fake transactions into the ledger.

\paragraph*{Autonomy.}
The main objective of BC is to switch the trust from one centralized authority to the asynchronous network of nodes~\cite{joshi2018survey}.
As discussed before, based on the consensus algorithm, every node can transfer and securely update data without any interference.

\paragraph*{Immutability.}
Records are immutably stored forever in blocks unless someone tries to alter them. Indeed, the consensus mechanism is theoretically vulnerable to attacks by miners attempting to use their hashing power to malicious ends.
Indeed, if a miner acting as an attacker or intruder can control the majority (i.e. $51\%$) of the total network mining power, he can attack the consensus mechanism so as to disrupt the security and availability of the BC. For instance, an attacker can cause previously confirmed blocks to be invalidated by forking below them and re-converging on an alternate chain~\cite{antonopoulos2017mastering}.
Actually, given the massive increase of total hashing power, this possibility is almost zero also for a pool of malicious miners.

\paragraph*{Anonymity.}
One of the key characteristics of BC is that it allows the users to perform pseudonymous transactions that are also verifiable and thus trusted.
Firstly, BC users are identified via a public address which does not contain any identifiable information to tie the address or the user.
Also, BC solves the trust issue from node to node. When a node transfers data or (crypto-)money toward another node, no one traces the origin and other nodes: all the details remain hidden and are recorded into the ledger.

\paragraph*{Open Source Implementation.}
Several BC open-source projects have been launched in the open-source community since BC inception.
Open source BC platforms enable autonomous developers to create decentralized applications based on BC technology which can be exploited by every interested user and are characterized by publicly verified records.

\subsubsection{Types of Blockchain.}
In the following, we discuss four different categories in which BC technologies can be divided based on their structural characteristics.

\begin{figure}[t]
    \centering
    \subfloat[][Public Blockchain.\label{fig:pbt}]{
    \includegraphics[width=0.3\textwidth]{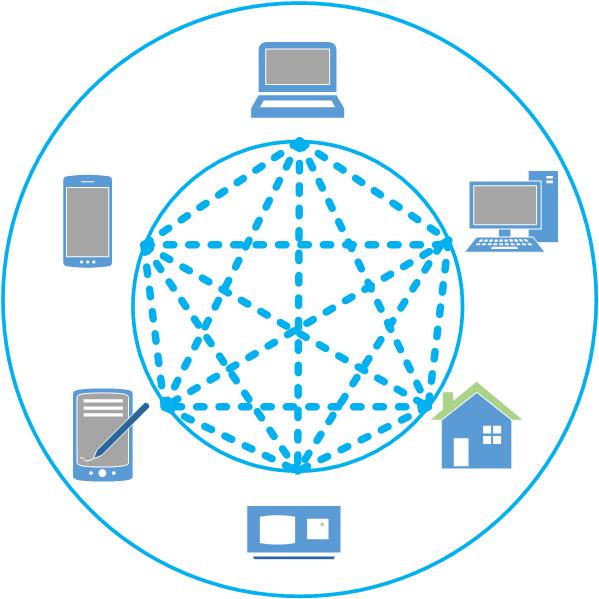}
    }
    \qquad
    \subfloat[][Private Blockchain.\label{fig:pbc}]{
    \includegraphics[width=0.3\textwidth]{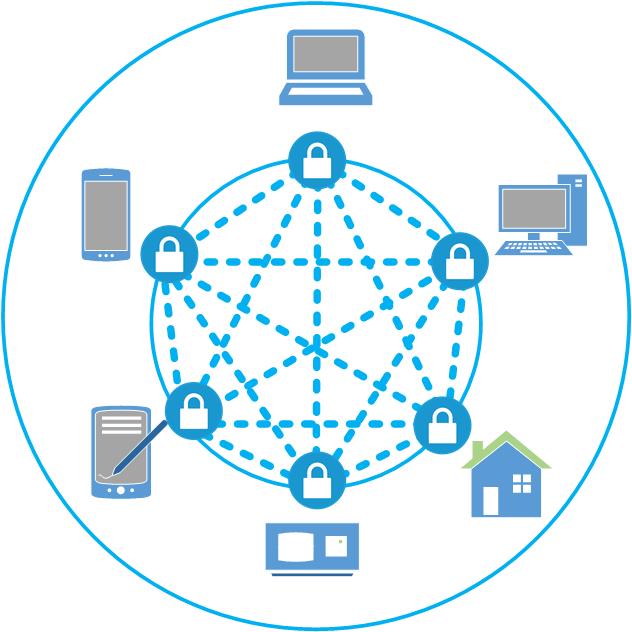}
    }
    
    \subfloat[][Consortium Blockchain.\label{fig:CB}]{
    \includegraphics[width=0.3\textwidth]{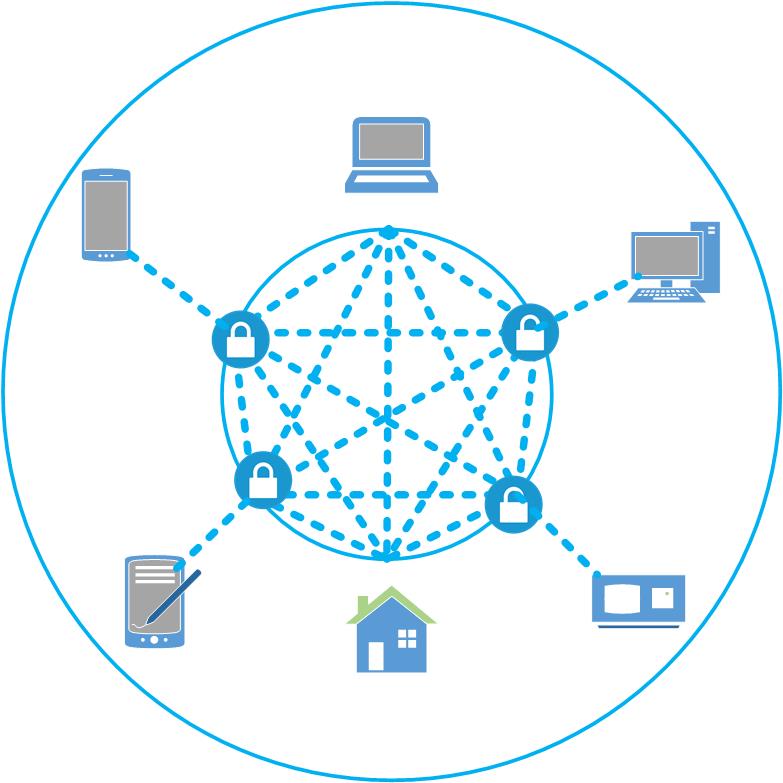}
    }
    \qquad
    \subfloat[][Hybrid Blockchain.\label{fig:ik}]{
    \includegraphics[width=0.3\columnwidth]{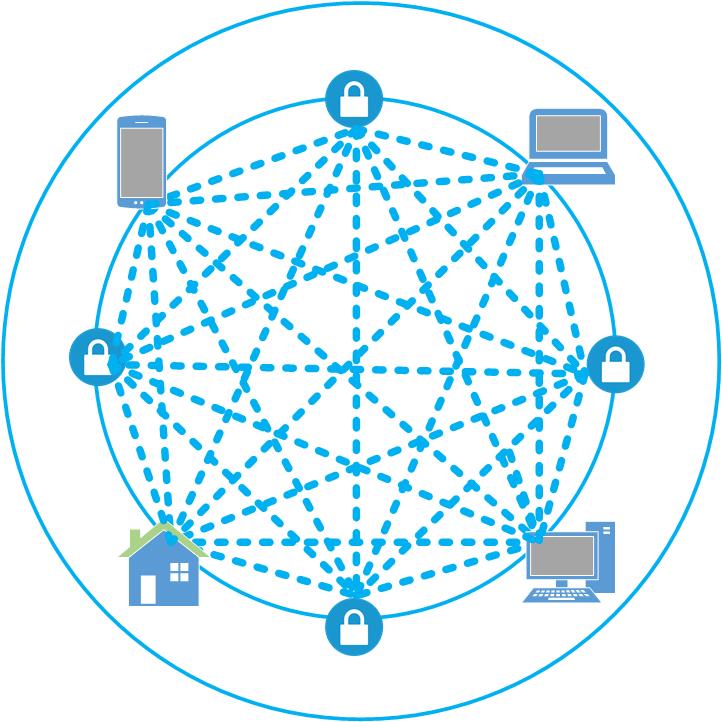}
    }
    \caption{Types of Blockchain.\label{fig:bc_types}}
\end{figure}

\paragraph*{Public Blockchain.}
A public BC is fully open source. Anyone can participate as a user, developer, or community member, without restrictions on new participants.
Figure~\ref{fig:pbt} sketches the structure of a public BC and shows the end users and their connections. All of them are acting as a block of the public ledger. The general user can have different nature and computational capabilities (e.g., a server or a smartphone user).
A public BC is also transparent because each member can independently access transaction details and control how the state of the BC evolves, and it is fully decentralized because no one centrally regulates the transactions.
Also, anyone can join the public BC network regardless of location or nationality, and the authorities can hardly shut down the accounts given their (pseudo-)anonymity.
Common examples of public BCs are Bitcoin, Ethereum, and Litecoin.

\paragraph*{Private Blockchain.}
A private BC is characterized by a single organization which has the sole control over the rules of the BC. Therefore, the participation is restricted by the authority which is in charge of managing and accessing the data.
Figure~\ref{fig:pbc} depicts the structure of a private BC.
The central web-like structure is the representation of the logical connectivity of the blocks.
Unlike the public one, the private BC has restrictions on the access of the transactions and related information.
Private BCs are indeed meant for a company who would like to collaborate and promote data but does not want that sensitive information is exposed on the public BC.
The entities operating in private BCs have full control of members and governance systems. Commonly, private BCs have a token associated with the chain and are used in supply chain management, digital identity, vote counting, asset ownership, hyperledger, etc.
Given its more centralized nature, for most applications, a private BC could replace a decentralized database.
Moreover, a private BC can be also used for educational purposes.

\paragraph*{Consortium Blockchain.}
A consortium BC has a selected group of participants (e.g., several organizations) called \emph{consortium} or \emph{federation}, which cooperate with the aim of taking advantage from BC capabilities (e.g., defining a system to reach consensus between organizations).
Figure~\ref{fig:CB} shows the general structure of a consortium BC. The major advantage of this type of BC over a private BC is that it depends on (and is managed by) several organizations instead of only one.
Users in the consortium can operate or run a node, can make transactions, and audit the BC. Commonly, participants can be individuals coming from banks, government, or supply chain management systems.
Common examples of consortium BCs are Hyperledger and Corda.

\paragraph*{Hybrid Blockchain.}
A hybrid BC tries to exploit the best aspects of the other types of BC (i.e. public, private, and consortium) to overcome their weaknesses and provide an efficient solution for trustworthy data sharing, access management, etc.
Figure~\ref{fig:ik} outlines the structure of a hybrid BC.
As shown in the figure, the blocks could be both restricted or public.
Thus, an authority can make the transaction ledger available to the users based on their needs.
With this type of BC, it is simpler to operate the business thanks to its functionalities that also preserve security and privacy. Indeed, it is more flexible and transparent for the business purpose to keep the data private and allows to decide what portion of data shall be made public. Nowadays, hybrid BCs are mainly used for data protection as in IoT networks and supply chain management systems.

\subsubsection{Attaining Consensus in Blockchain.}
\label{subsec:consensus}
As discussed above, consensus algorithms are of paramount importance in a BC: they are used to decide how a new block is verified and added to the chain.
Hereinafter, we discuss two common types of consensus algorithm: proof-based and voting-based~\cite{nguyen2018survey}.

\paragraph*{Proof of Work (Pow).}
PoW is a proof-based consensus algorithm in which miners need to prove that they have done a certain amount of computing work to be allowed to publish.
In detail, the PoW is included in a new block and acts as proof that the miner spent significant computing effort.
PoW requires to solve a complex mathematical problem~\cite{gervais2016security} based on a hash-cryptography puzzle.
The latter needs significant computational power to be solved but its solution can be quickly verified.
Miners compete each other for accomplishing this task, and when a miner obtains a solution, the latter is broadcast to the network. All other miners then verify if the solution is correct, and in this case the new block is added to the chain.
The competition to solve the PoW algorithm to earn reward and the right to record transactions on the BC is the basis for BC security model~\cite{antonopoulos2017mastering}.
For instance, PoW can be used to prevent cyber-attacks such as Distributed Denial-of-Service (DDoS) that would be inefficient since the cost incurred for exhausting BC resources via the DDoS would be significantly greater than the potential rewards for attacking the network.

\paragraph*{Proof of Stake (PoS).}
PoS is a proof-based consensus algorithm mainly used in public BC, which aims to overcome the unfairness of PoW. Indeed, the latter favors miners having more powerful equipment, which can thus find a suitable solution to the cryptography puzzle easier than other miners with less powerful equipment.
More specifically, PoS is based on the idea that a miner who
owns much stake (i.e. the percentage of coins held) would be more trustful.
PoS is considered as less risky in terms of revenue for miners to attack the network, as it structures compensation in a way that makes an attack less advantageous for a malicious miner.

\paragraph*{Voting-based Consensus.}
Voting-based consensus algorithms presume that the nodes should be known, in contrast to proof-based algorithms where nodes can freely join or leave the network.
Therefore, they are more suited to private or consortium BCs.
Furthermore, the nodes have to both maintain the ledger (as in proof-based) and jointly verify new blocks via a voting mechanism (differently than proof-based).
Each node communicates with other nodes to decide if adding the new block to the chain: a certain number of nodes (depending on the specific voting mechanism) have to verify the same proposed block before validation.

\subsection{Software Defined Networking}
\label{subsec:sdn}

\begin{figure}[t]
    \centering
    \includegraphics[width=0.8\textwidth]{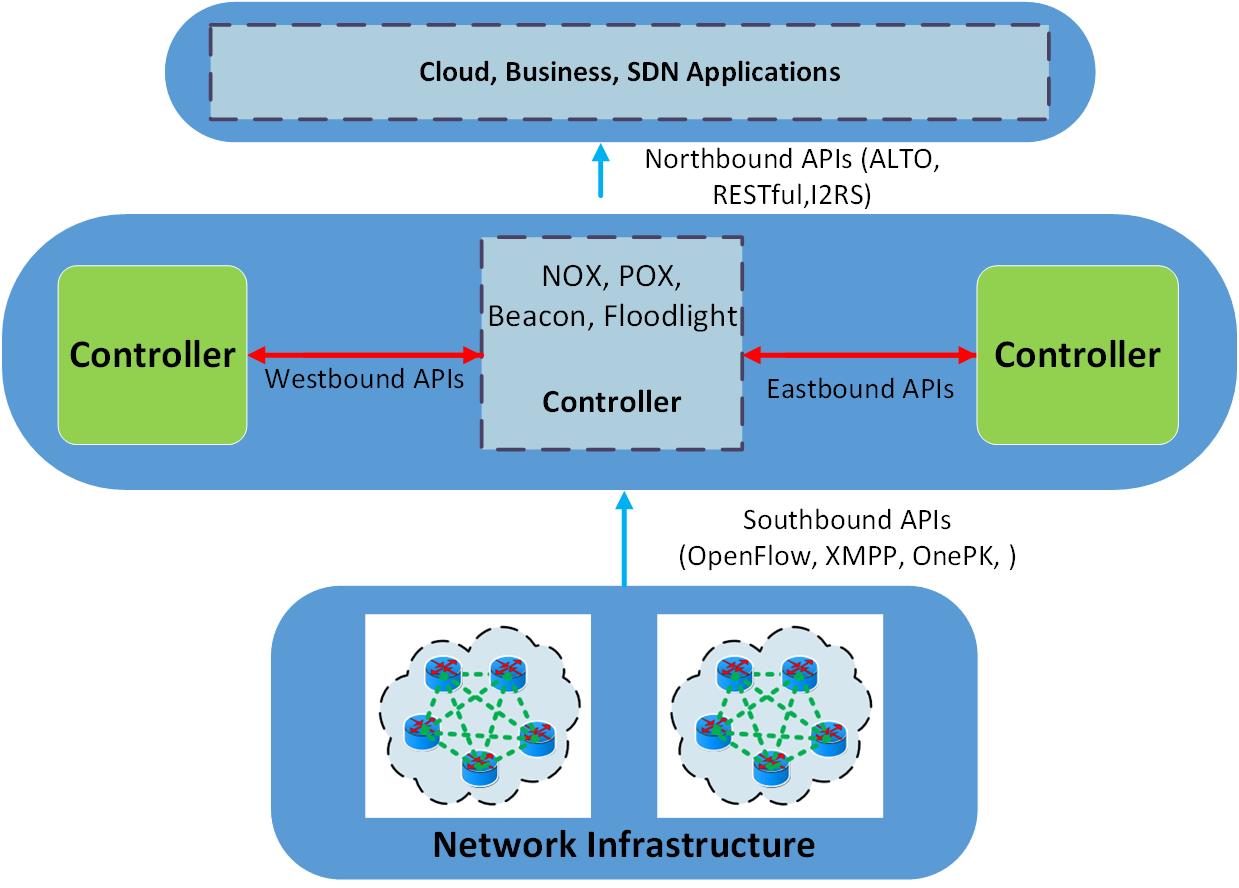}
    \caption{General SDN architecture.}
    \label{fig:sdn}
\end{figure}

In conventional systems, network connections are established via switches and routers, which are also responsible for transmitting data over the network. Such a networking scheme may be subject to lack of confidentiality and then could be prone to third-party attacks. For these reasons, the development of the SDN paradigm is crucial. SDN is a networking strategy that efficiently provides the facility of a centralized environment and aims to decouple data transfer process from the devices designed for it~\cite{kreutz2014software}. This paradigm is based on the existence of different \textit{planes}, each responsible for some specific functions: (i) the data plane is responsible for packet forwarding, (ii) the control plane decides on routing using a flow table containing rules to properly manage incoming packets, and (iii) the application plane encompasses the services made available to the users.
It is worth noting that flow rules can be easily changed according to the needs~\cite{ali2015survey} by exploiting the improved programmability and control provided by SDN compared to conventional networking systems~\cite{DBLP:journals/comcom/MegyesiBAPM17}.
OpenFlow is the most known protocol used for the communication between the switches and the central controller in SDN environments (despite it still has some
vulnerabilities~\cite{9194211}).
Figure~\ref{fig:sdn} depicts the structure of a common SDN architecture, whose planes and related interconnections are described hereinafter.

\paragraph{Data Plane.}
The data plane (also known as edge or infrastructure plane as reported in Fig.~\ref{fig:sdn}) is the lowest plane of an SDN architecture and comprises the devices used for data forwarding such as switches (physical or virtual), access points, routers, etc.
The communication between the data plane and the control plane is established with the OpenFlow protocol through the \textit{Southbound Application Programming Interface (API)} commonly using encrypted channels.
The data plane performs packet forwarding based on the flow rules providing forwarding logic, which are defined according to the OpenFlow specification and installed through the Southbound API by the control plane~\cite{shaghaghi2020software}.
Flow rules include MAC and IP destination addresses, transport-layer source and destination ports, and other necessary information for matching the desired packets.
Notably, OpenFlow rules can be exploited to implement firewalls and enforce various security policies.

\paragraph{Control Plane.}
The control plane is the middle (viz. backbone) plane in an SDN architecture. It is considered as the brain of the networking system~\cite{ali2015survey} and is in charge of managing the routing process.
In more detail, the control plane encompasses logic and functional controllers applied to manage the control logic at different levels and to provide controlling functionalities, respectively. As shown in Fig.~\ref{fig:sdn}, its core can be realized via different OpenFlow-compliant implementations (e.g., NOX, POX, Beacon, Floodlight, etc.)~\cite{karakus2017survey}.
It is connected with the application plane through the \textit{Northbound API} and with data plane through the \textit{Southbound API}.
Additionally, two other interfaces are the \textit{Eastbound API} and the \textit{Westbound API} used for the communication between multiple distributed controllers~\cite{abuarqoub2020review}.
It is worth noticing that the control plane provides several enhanced network services which offer improved management, QoS, and data privacy and security to the network infrastructure.

\paragraph{Application Plane.}
The application plane is the topmost plane of an SDN architecture. It includes programs to control the networking system and establishes a connection with the control plane using the \textit{Northbound API} (usually a RESTful API).
It also provides higher-level services to the SDN users and applications running on top of existing controller platforms. In detail, from this plane, applications transmit their requirements to the SDN controller where they are processed and then translated into commands and rules for the data plane devices of the SDN architecture. Some of the most common SDN applications are routing, load balancing, and firewalling~\cite{bannour2017distributed}.

\section{Integration of Blockchain with SDN}
\label{sec:integration_sdn_bc}

\begin{figure}[t]
   \centering
   \includegraphics[width=0.7\textwidth]{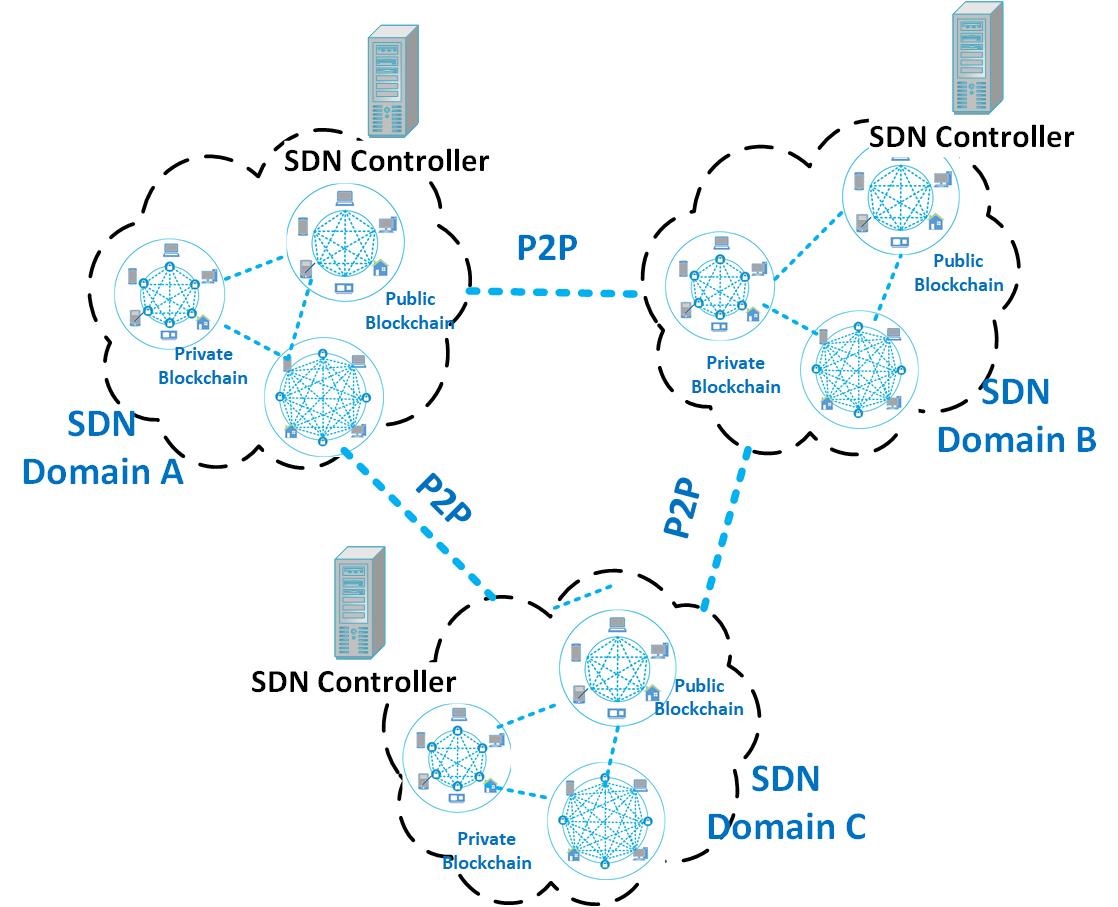}
   \caption{Example of the integration of Blockchain with SDN.}
    \label{fig:ssbs}
\end{figure}

This section describes the integration of SDN and BC technologies, whose fundamentals have been provided in Sec.~\ref{sec:overview}.
Figure~\ref{fig:ssbs} depicts a common scenario in which both public and private BCs are integrated in a multi-domain network defined via the SDN paradigm. Each domain uses a distinct SDN controller, which communicates with the others via a P2P connection.
Specifically, Sec.~\ref{subsec:sdn_bc_motivations} reports reasons why researchers have proposed this integration and Sec.~\ref{subsec:bc_sdn} collects state-of-art proposals.
We point the interested reader to the specific works reviewed hereinafter, each proposing a particular realization of the general scenario exemplified in Fig.~\ref{fig:ssbs} to fulfill the different needs (e.g., reducing latency, optimizing resource usage, guaranteeing security and privacy) of considered application scenarios.

\subsection{Motivations}
\label{subsec:sdn_bc_motivations}
In modern networks (e.g., always online smartphones, IoT sensors), the enormous number of interconnected devices produces a considerable amount of data gathered from the actual world. In this scenario, a urgent need to properly manage data arises and SDN is proposed as a solution because it can programmatically handle all the data received from the network environment~\cite{cui2016big}.
Altering the base architecture of a network system is a challenging task; SDN makes it much easier and allows to quickly modify network characteristics.
Researchers are prompted to take advantage of the capabilities of the SDN paradigm that can be exploited to performs distinct operations on the data based on the specific layer on which they operate~\cite{rahman2020distblockbuilding}.

Unfortunately, the communication among SDN planes is not fully secured and hence its adoption could lead to a corrupted network system due to maliciously wrong information exchanged~\cite{pritchard2017security}.
To enhance the security and reliability of SDN networks, BC could efficiently help in checking the authenticity of transmissions~\cite{ourad2018using,hammi2018bubbles,lin2018bsein}.
Moreover, the addition of Smart Contracts (SCs) into the BC validation process makes this technology even more trustworthy~\cite{gatteschi2018blockchain}.
Besides, BC keeps the history of transaction in an immutable ledger that can prevent any third party interaction and tampering.
However, on the one hand, BC can take the responsibility of the security of data transmission but on the other hand, it can not manage the data from IoT directly.
Therefore, the integration of SDN and BC has been pursued to jointly take advantage of the peculiarities of both technologies.

\subsection{State-of-Art Proposals}
\label{subsec:bc_sdn}
We now discuss the research works that have proposed the joint usage of BC and SDN. Many researchers have suggested different BC-SDN architectures to guarantee latency, network performance, resource usage, security, and other desirable features~\cite{sharma2017distblocknet,el2019cochain,muthanna2019secure,zhang2019blockchain,navid2019SDIoBoT,derhab2019blockchain,sharma2017software}. In the following, we go into details of such works.

Xie et al.~\cite{xie2019blockchain} leverage SDN for 5G VANETs and show that their BC-based IoT network can unveil malicious vehicular nodes and messages.
Similarly, Zhang et al.~\cite{zhang2019blockchain} also introduce a distributed BC-based software-defined VANET framework for smart cities named \textit{Block-SDV}.
In detail, Block-SDV integrates a novel deep Q-learning approach to solve the optimization problem of finding the optimal features for BC (e.g., trust features of BC nodes, number of consensus nodes, trust features of each vehicle, and computational capability of BC).
The effectiveness of Block-SDV and related deep Q-learning is demonstrated in a simulation environment.
Gao et al.~\cite{gao2019blockchain} present a trust-based model to detect malicious activities integrating BC and SDN to improve security in 5G and fog VANET networks.
The joint adoption of BC and SDN technologies have been also exploited for the design of Intrusion Detection Systems (IDSs) employed in the control framework of industries~\cite{derhab2019blockchain}.
Houda et al.~\cite{el2019cochain} propose a BC-based architecture named \textit{Cochain-SC}, having two levels of attack mitigation: intra-domain and inter-domain DDoS mitigation in SDN.
More specifically, the combination of intra entropy-based, intra Bayes-based, and intra-domain mitigation schemes is used to classify and mitigate the impact of malicious traffic. Furthermore, for inter-domain, the authors introduced an SC-based architecture that considered Ethereum technology to simplify the collaboration among SDN-based Autonomous Systems against DDoS attacks.
Finally, the performance of their proposal is assessed based on efficiency, flexibility, security, and cost effectiveness.

Sharma et al.~\cite{sharma2017distblocknet} propose the \textit{DistBlockNet} architecture that allows the cooperation of SDN and BC in an IoT network.
The authors devise a strategy for updating and formalizing the flow rule table applying a BC, and evaluate the performance according to various metrics showing better results compared to previous works.
Rahman et al.~\cite{rahman2020distblockbuilding} present \textit{DistBlockBuilding}, a distributed BC-SDN architecture for smart cities, designing also a cluster-head selection algorithm for collecting sensors data with low energy dissipation. The authors evaluate the performance with different parameters such as throughput, latency, and packet arrival rate.
Chaudhary et al.~\cite{chaudhary2019best} leverage the benefits of BC technologies for the transportation network and use SDN to provide low power consumption and legitimate resources.
Secure and reliable energy management is also considered in~\cite{zhiyi2018cyber} where the authors design a cyber-secure decentralized framework to ensure reliability, efficiency, and sustainability jointly using SDN and BC.

With a specific focus on fog environments, Muthanna et al.~\cite{muthanna2019secure} introduce an IoT-based fog system that incorporates BC and SDN.
In detail, SDN can attain high privacy, availability, and security for IoT-based applications, while BC assures decentralization in a secure way.
Furthermore, they investigate latency, network efficiency, and resource utilization for the experimental evaluation of their architecture.
Similarly, Sharma et al.~\cite{sharma2017software} present a novel BC-based distributed cloud architecture with fog nodes acting as SDN controller at the edge of the network, thus proposing an innovative combination of fog computing, SDN, and BC, along with an architecture for supporting availability, real-time data bringing, scalability, security, resilience, and low latency.
Using this architecture, they also evaluate throughput, response time, and accuracy in detecting real-time attacks.

IoT networks is considered in~\cite{navid2019SDIoBoT}, where authors present a model for facing present IoT challenges with SDN and BC in the context of 5G networks.
They also introduce the Elliptic Curve Digital Signature algorithm for security purposes which is evaluated in a simulation environment.
The challenges for securing cloud management from network perspective are investigated in~\cite{9045378}, whose authors
take advantage of SDN and BC to perform secure cloud management.
An infrastructure incorporating both these technologies is developed to improve the availability of cloud systems through the automatic provisioning of network bandwidth.
Derhab et al.~\cite{derhab2019blockchain} also face security challenges in an IoT-based industrial environment via a control framework leveraging BC-SDN integration.
Moreover, Open-Flow based flow rules are installed in this industrial IoT network effectively exploiting the SDN paradigm.
The presented SD-WAN model admits two additional elements, namely: (i) an IDS which can defend against attacks using the joint combination of Random Subspace Learning and K-Nearest Neighbors methods (\textit{RSL-KNN}), and (ii) a BC-based integrity checking system.

Differently, Qiu et al.~\cite{qiu2018permissioned} deal with the consensus problem in BC for Software Defined Industrial IoT.
To address such a problem, the authors use a Q-learning approach showing also its effectiveness via simulations.
Shao et al.~\cite{shao2019blockchain} also present a consensus algorithm named \textit{Simplified Practical Byzantine Fault Tolerance (SPBFT)} to transmit messages in the SDN network securely.
Specifically, the authors apply this BC approach in an SDN environment to create a readable, addable, and unmodifiable decentralized database. Also, to enhance the security and reach consensus in the SDN control layer, they leverage the proposed SPBFT algorithm to transfer messages between controllers. Finally, they compare SPBFT with the Practical Byzantine Fault Tolerance algorithm proving that the proposed algorithm can significantly improve performance in terms of security and efficiency.
On the same line, Liu et al.~\cite{liu2019trust} define the \textit{TrustBlock} method to calculate the trust value of SDN network node based on BC.
More in detail, the authors leverage SDN to assess the legitimacy of the nodes and BC to ensure the trust value authenticity, immutability, and openness. The weight of each evaluation attribute is determined using an entropy-based method.
BC is also considered as a solution to improve the SDN security of IoT-based applications interacting via inter-cloud communication in~\cite{tselios2017enhancing}.
Basnet et al.~\cite{basnet2017bss} propose \textit{Blockchain Security over SDN (BSS)} to protect the privacy and availability of resources against non-trusting members. The authors demonstrate that BSS facilitates files sharing among SDN users in distributed P2P basis using OpenStack as a cloud storage platform.
Additionally, Kataoka et al.~\cite{kataoka2018trust},
propose a Trust List that represents the distribution of trust among IoT-related stakeholders and aims to verifying and trusting IoT services and devices avoiding undesirable traffic of IoT devices responsible for attacks on the network system.
Firstly, they verify that IoT traffic management is properly achieved using their proposal.
Then, they discuss and implement a proper combination of BC and SDN to ensure security, dependability, and trusting for IoT services and devices.
Finally, the proof of concept open-source implementation is tested on both public and private BCs.

Steichen et al.~\cite{steichen2017chainguard} propose \textit{ChainGuard}, which utilizes an SDN module to manage network collisions and implements a firewall based on BC. Furthermore, ChainGuard can offer access control capabilities and can effectively mitigate flooding attacks. 
Finally, the authors discuss other aspects of the proposed architecture and provide observations on their experiments with the aforementioned model.
Similarly, Houda et al.~\cite{abou2019co} present framework sharing SDN and BC to mitigate DDoS attacks in a scalable, stable, and cost-effective manner.
Another possible application is proposed in~\cite{okon2020blockchain} whose authors introduce a consolidated BC-SDN system that manages spectrum assets to enable interoperability between mobile network operators over small cells.

\section{Security and Privacy Issues in BC-SDN Systems}
\label{sec:security_privacy}
This section covers security and privacy issues concerning BC and SDN technologies and their integration. Precisely, in Sec.~\ref{subsec:bc_security}, we discuss issues in BC, also addressing those related to scalability and confidentiality.
Section~\ref{subsec:sdn_security} deals with security management in SDN with particular focus on IDS.
Section~\ref{subsec:bc_sdn_security} describes issues arising when integrating BC with SDN and also reports new challenges faced in this context.

\subsection{Security and Privacy Issues in Blockchain}
\label{subsec:bc_security}

As described in Sec.~\ref{subsec:blockchain}, BC is widely used in different domains that take advantage of its unique properties.
Nevertheless, dangerous attacks can cause several issues in BC-based networks.
In the following, we describe some security issues that attackers exploit to undermine the safety and cause damage in BC systems~\cite{wang2018overview}.

\subsubsection{Majority Attacks ($51\%$ Attack).}
When BC uses PoW to reach consensus, the probability of mining blocks depends on work done by miners. If a miner holds $51\%$ (i.e. more than half) of total computing power, it can monopolize the mining process and take control over the BC to decide which block is validated and then added to chain~\cite{wang2018overview}.
Based on this reasons, miners could join together to mine more blocks, thus constituting a \enquote{mining pool}, that is a group of nodes holding most of computing power.
In such a situation, miners (organized in a mining pool) can modify the transaction data (potentially causing double-spending attacks), alter their order, and stop the mining of other available blocks~\cite{lin2017survey}.

\subsubsection{Forking Attacks.}
In a BC system, users can propose changes of BC protocol or software. When this situation occurs, miners have to decide which version to use and if there is not a unanimous decision, two BC versions are created. Therefore, nodes can be divided into two types: old and new nodes.

Forks are categorized into two types: \textit{hard-fork} and \textit{soft-fork}.
\begin{itemize}
    \item Hard fork: a hard-fork is a significant change within a cryptocurrency protocol that is incompatible with the previous version. It usually changes or improves an existing protocol or creates a new independent protocol (and consequently a new chain). The node that does not update to the new version will not be able to process transactions or push new blocks to the BC. If a group of nodes continues to use the old version, a permanent split occurs in the BC. 
    
    \item Soft fork: a soft-fork is actually a change in a cryptocurrency protocol that is back-compatible and where non-updated nodes are able to process transactions and add new blocks to the BC. 
\end{itemize}
With forking attacks, malicious users can replace the longest chain (also referred to as the most trusted chain) on the current network with another chain to gain personal benefits.
Therefore, reducing the generation of fork to the minimum necessary is the most directly method to defend against forking attacks in a BC system.

\subsubsection{Selfish Mining.}
Selfish mining constitutes another significant concern for BCs. During this attack, the malicious miner maintains the mined blocks without broadcasting them towards the networking system and generates a personal chain where he is the only one producing blocks that are revealed only after particular requirements (chosen by the attacker) are fulfilled.
In other words, the attacker generates a secret longer chain that when revealed would have more ``chain-work'' (viz. PoW) than the public shorter chain, and thus will be followed by other miners cancelling the revenues obtained on the public chain in the meantime.
Therefore, in this case, reliable miners spend lots of time and resources without obtaining any revenue, while selfish miners continue to mine their individual chains obtaining high revenue after revealing them~\cite{joshi2018survey}.

\subsubsection{Social Engineering.}
Social Engineering is an attack conducted by outsiders using psychological tricks to get users' personal information to access a computer or even a whole network~\cite{peltier2006social}.
Social Engineering is a main issue afflicting BC security: according to~\cite{ledgerops}, in $2018$ almost $\$3$ billion was lost due to social engineering. 
To perform this attack, phishing is one of the most used technique. The attacker sends the targeted users a fake URL or website links on the behalf of a company or organization brand he trusts. Commonly, they pretend that his account has some security issues and ask the users to send their information through provided links to solve such issues, claiming that otherwise his account will be blocked.
The ultimate aim of the attacker is to put pressure on the user and hands over his information.
Social engineering attacks are particularly target toward crypto users and realized via phishing among which SIM swapping is one of the most utilized.

\subsubsection{Privacy in Blockchain: Confidentiality and Authentication.}

\begin{figure}[t]
    \centering
    \includegraphics[width=0.6\textwidth]{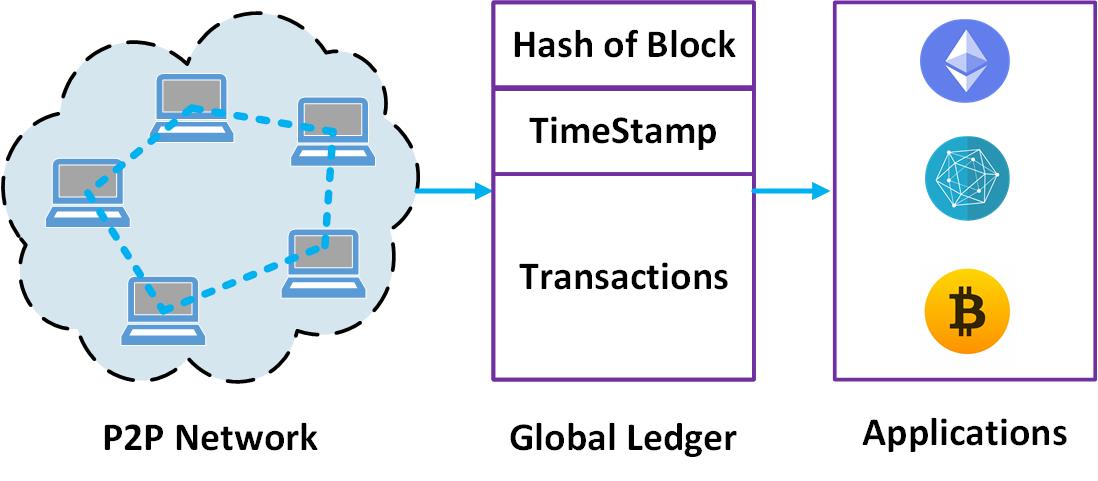}
    \caption{Overview of confidentiality and authentication process of a Blockchain.}
    \label{fig:pbcp}
\end{figure}

A BC is a decentralized and distributed technology composed of a huge number of interconnected blocks which contain key information generated by different parties mainly for business purposes~\cite{rahman2021smartblock}.
Researchers are attracted by the properties of BC and have proposed it for IoT, Industry 4.0, and many other environments to increase security and privacy.

As described, BC uses consensus mechanisms such as PoS and PoW being independent of any third party: confidentiality and secrecy of transactions are guaranteed since they are not verified and owned by a single entity. Another technique named \emph{Zero-Knowledge Proof} transmits the information to an examiner to identify whether the transaction is valid or not~\cite{li2020privacy}. The goal of such a technique is to hide the internals of a transaction revealing just its validity and thus guaranteeing the full confidentiality of the transaction details.

Moreover, to preserve the privacy (i.e. to further guarantee their confidentiality), the transactions are secured via \emph{public-key cryptography}, and once a block of information is included in the chain, it cannot be altered or modified~\cite{yu2018blockchain}.
Public-key cryptography is also exploited for digital signature used for transaction authentication in an untrustworthy environment~\cite{feng2019survey}. 
In detail, the BC leverages public-key cryptography to send transactions and verify their authentication.
The transaction is signed using the sender's private key, before being sent over the P2P network (see Fig.~\ref{fig:pbcp}). To this aim, existing BC typically employs the elliptic curve digital signature algorithm.
As shown in Fig.~\ref{fig:pbcp}, P2P networks with the aid of global ledger give users the complete control on their data to ensure confidentiality and authentication to different applications, decreasing the threat of third parties to sell, store, or manipulate personal information.
Typical applications needing such privacy requirements are financial transactions, healthcare record, and legal documents (cf. Sec.~\ref{sec:applications}).

Furthermore, BC is capable of providing privacy by storing data transactions as blocks with hash values.
The blocks are organized as a mesh like structure and are encrypted and decrypted via public and private keys, respectively~\cite{mohanta2020addressing}. During this process users exploit the hash value of the public key as their address and can also generate several addresses. This in turn keeps the real identity of a user.
In such a way, data confidentiality is preserved, and each transaction can be checked with the aid of the consensus process and the logs inside the BC.
Indeed, using this procedure, every transaction is rechecked until it is globally included: it is first signed by the recipient and then added with a digital signature.
Overall, all these procedures help to guarantee the integrity of data transferred within the network and maintain the most sensitive data confidential~\cite{zhang2019security}.

\subsubsection{Authorization.}
When data reside in and are managed by a single organization, dealing with security and privacy issues is relatively simple.
Conversely, when the information is exchanged between different domains---as when leveraging BC---securing data is a much more complex process.
Secure access control mechanism is a common approach used to ensure that only authorized entities can access shared data.
Such a mechanism involves access policies commonly consisting of Access Control Lists (ACLs) associated with the data owner. An ACL is a list of requestors who can access data, and provides related permissions (i.e. read, write, update, delete) to specific data.
Hence, the authorization is a function of granting permission to authenticated users to access the protected resources following predefined access rules.
Access rules mainly focus on who is performing which action on what data object and for which purposes.

Commonly, traditional authorization approaches are deployed, managed, and run by third parties (e.g., cloud service providers) that can be benign but curious.
To build a trustworthy system, the BC can be combined with an access control mechanism, realizing a self-management of users' own data, and thus keeping shared data private.
For instance, BC users can leverage SCs (cf. Sec.~\ref{sec:applications}) to define access permissions (i.e. authorize, refuse, revoke), operations (i.e. read, write, update, delete), and duration of their data sharing without the loss of control right.
SCs can be triggered on the BC once all the preconditions are met and can provide an audit mechanism for any request recorded in the ledger as well.
A notable application of BC-based authorization leveraging SCs is for securing healthcare data sharing~\cite{shi2020applications}.

\subsubsection{Other Issues.}
Given the rising number of transactions, the BC volume grows on a daily basis, and scalability issues have to be taken into account.
Indeed, every single node must save every transaction to verify all of them when needed.
For instance, Bitcoin---the most common crypto-currency using BC---has to process $\approx7$ transactions per second~\cite{zheng2017overview}.
Given the small size of a block capacity, Bitcoin can not handle or process the millions of transactions in real-time, and many small transactions are delayed. Consequently, miners request high fees to process and validate them.

Additionally, confirmation time is another challenge to consider when operating with decentralized P2P transactions as in BCs.
More specifically, each transaction takes an average time between $20$ and $40$ minutes for confirmation~\cite{reyna2018blockchain}.
Optimization strategies based on division of BC nodes, taking into account the number of computers the user has to access the network, have been proposed to reduce confirmation time up to more than $70\%$~\cite{pazmino2015simply}.

\subsection{Security Management in SDN}
\label{subsec:sdn_security}
\begin{figure}[t]
    \centering
    \includegraphics[width=0.6\textwidth]{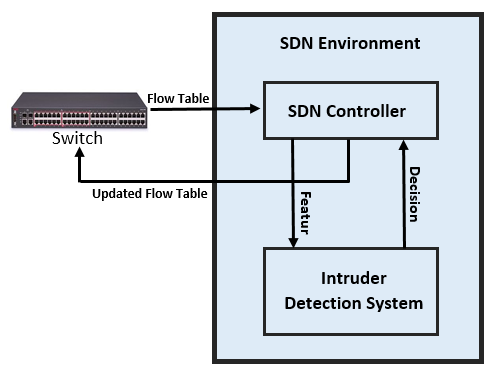}
   \caption{Block diagram of an IDS.}
    \label{fig:IDSprocess}
\end{figure}

As shown in Sec.~\ref{subsec:sdn}, SDN is a scalable networking paradigm consisting of three planes: the application, control, and data plane.
To secure SDN and lead to adequate protection, the first step is to understand the weaknesses of each of them.
Indeed, it is worth noticing that the security of the overall system can be achieved by securing all the above-mentioned planes.
However, because such planes are separated~\cite{rahman2021smartblock}, the protection of the whole system needs also to be assured. In this sense, to secure the SDN model from attacks and intrusions, an additional plane can be added: it comprises attack and intrusion detection systems, and a firewall to defend the network and guarantee its availability~\cite{zkik2019design}.

Generally, in an SDN, users access data from various controllers, thus implementing a new architecture where they ask permission from the controller may protect the system from malware. On the other hand, networking devices (e.g., switches, routers, storage systems, and sensors) are involved in packet forwarding and located in the data plane. In addition, specific tools, such as \textit{Flow Checker}, \textit{VeriFlow}, and \textit{FortNOX} can be used to effectively manage the packets by systematically forwarding or deleting packets thus increasing network efficiency~\cite{shaghaghi2020software}.
OpenFlow can assist devices in determining the exact match of the packets~\cite{chica2020security}.
While attack detection protects the network from the intruders, data encryption handles users authentication and protects the data from being manipulated by the attackers.
In addition, the state of the network can be determined by analyzing the information processed by the control plane.
In the last years, different tools have been proposed to support network administrators like \textit{Athena}~\cite{lee2017athena} that performs anomaly detection via a scalable approach.

As depicted in Fig.~\ref{fig:sdn}, in an SDN environment, there can be multiple controllers (communicating via the eastbound/westbound API) or a single centralized controller.
Data transmission is performed by the switches following a flow table that contains information about the connections and their properties~\cite{Islam_Rahman_Kabir_Khatun_Pritom_Chowdhury_2021}.
These devices take specific decisions for each request coming from different connections.
In this context, a malicious user can shake up the normal traffic flows in the network.
For example, an intruder may send different request for the same service to damage the network or sends a message that corrupts network devices.
These are examples of common situations in which an IDS can help to ensure security~\cite{shamshirband2020computational}. Figure~\ref{fig:IDSprocess} depicts the main components of an IDS with the principal information exchanged among its blocks when applied in an SDN. Based on the information exchanged with the controller(s), the IDS can guide the update of the flow table to avoid malicious threats conducted by an attacker.
To attain this goal, some researchers propose to check for the entropy of the IP addresses to discover anomalies regarding the users, and then analyze the flow table to figure out some useful features to build Machine Learning models for future predictions on unseen data~\cite{yang2018ddos,Islam2021}.

\subsection{Security and Privacy Issues of BC-SDN}
\label{subsec:bc_sdn_security}
The combination of BC and SDN helps to improve the efficiency of smart architectures such as: smart grid, education, healthcare, industry, transportation, etc. Nevertheless, security continues to be a severe concern for these systems, and efforts are needed to solve challenges and issues as described in the following.

\subsubsection{Aim and Challenges of BC-SDN for Security.}

The SDN-based system separates the control plane from the data plane to provide more flexibility to the network. However, the control plane is the main target of the intruders who can easily take control over it.
Therefore, BC-SDN aims to improve the overall security and flexibility of the networking system~\cite{shao2019blockchain}.
BC is compatible with both private and public environments and it can be useful for securing P2P communications.
For transferring file securely in IoT networks, BC provides security, while SDN is also beneficial for reducing energy consumption~\cite{khattak2020dynamic}.
For these reasons, we can identify many areas where there is a need to implement the integration of SDN with BC.

However, several challenges concerning this combination still remain, and the systems where the joint combination of the two techniques is applied have to deal with some issues.
Firstly, the planes of SDN networks can be accessible from the intruders who can obtain information such as host IP or the networking architecture performing simple scanning operations~\cite{thangavel2019network}.
Leveraging these data attackers can manipulate packets and launch several attacks such as \textit{ARP} or \textit{IP spoofing}~\cite{priya2021mitigation}.
Moreover, if a host network is attacked then the intruder can easily take control and manipulate the communication from the control plane carrying out a \textit{hijacking} attack~\cite{gadze2021investigation}.
A \textit{man in the middle} attack instead aims to take control over the SDN flow tables and rewrite their content.

BC has been proposed as a viable way to protect the SDN environment but actually it is not a silver bullet for all SDN security concerns and some challenges remain open.
Among possible attacks, DDoS is one of the most dangerous one~\cite{altarawneh2021availability} since it impairs the availability of the chain and can allow attackers access to associated wallets and exchanges.
Also, the presence of unauthorized users might create security issues such as malicious access to wallet and refuse to allow new blocks from entering the BC.
Moreover, in the BC-based system, transactions data are usually store on public chain so that anyone can easily get access to the records.
This would create important issues for environment where transactions are meant to be secret (e.g., bank transactions).
Time and cost are another two challenges of integrating BC with SDN. Since smart systems need to deal with several quick-access tasks, the lesser the time, the higher the efficiency.

\subsubsection{Confidentiality and Privacy in BC-SDN.}
Along with security, privacy is a key concern also when considering the integration of BC with SDN.
Indeed, the combination of such technologies can significantly help in reaching this goal.
As discussed in Sec.~\ref{subsec:bc_security}, BC stores information in the blocks after performing cryptography and data hashing, thus privacy is guaranteed, and personal information could remain confidential.
Since centralized systems can be affected by single point of failure~\cite{DBLP:journals/jnca/AcetoBMPP18}, decentralized structure of BC comes up to reduce this vulnerabilities~\cite{yu2018blockchain}.
In SDN, data privacy could be undermined by the collection of the information since raw data are gathered from different types of device and processing procedures can corrupt data integrity.
Considering a real use case, the deployment of closed circuit television cameras for security purposes is becoming extremely common in smart environments, and as a result, people can barely attain a complete privacy because they feel monitored all the time with consequent discomfort.
BC could be applied for privacy protection so that actual records are shown only to authorized users; otherwise the videos are blurred by the system~\cite{fitwi2019lightweight}.

\section{Notable Applications}
\label{sec:applications}

This section describes applications developed thanks to facilities and properties offered by studied technologies. Section~\ref{subsec:bc_applications} provides the description of the most important apps supported by BC technology; the last section discusses the ones built by integrating SDN with BC (Sec.~\ref{subsec:bc_sdn_applications}).

\subsection{Blockchain Applications}
\label{subsec:bc_applications}
\begin{figure}[t]
    \centering
    \includegraphics[width=0.55\textwidth]{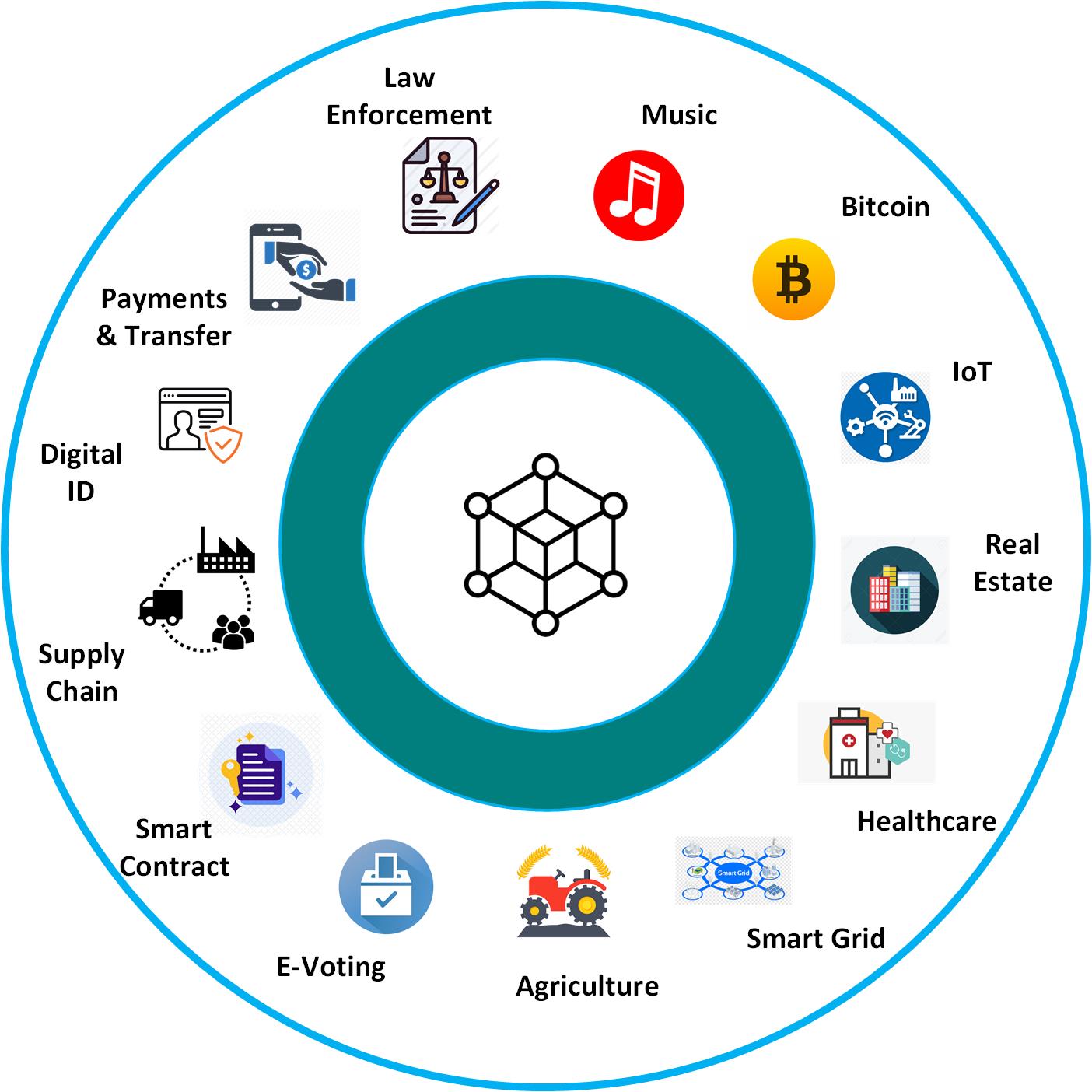}
    \caption{Blockchain applications.}
    \label{fig:bcapp}
\end{figure}

\begin{table*}[t]
\scriptsize
\caption{State-of-the-art works focusing on the utilization of Blockchain in different application fields. The works are reported in chronological order.}
\label{tab:BC-App}
\centering
\resizebox{\textwidth}{!}{
\begin{tabular}{@{}lll@{}}
\toprule
\textbf{Ref.} & \textbf{Year} & \textbf{Application Field} \\
\midrule
\cite{li2021blockchain} & 2021 & 
Application of BC for the sustainability of Prefabricated Housing Construction.
\\
\midrule

\cite{kouhizadeh2021blockchain} & 2021 & 
Barriers for adopting BC technology in sustainable supply chains. 
\\
\midrule

\cite{9499121} & 2021 & Constructing a smart city IoT framework with BC and SDN that is energy aware and distributed securely. 
\\
\midrule
\cite{hamledari2021application}& 2021 & Blockchain application for supply chain in the area of construction industry.\\
\midrule

\cite{hong2021public} & 2021 & Blokchain technology for the safety of the management of food data.\\
\midrule
\cite{mishra2021application} & 2021 & Application of Blockchain for financial sector.\\
\midrule

\cite{9641303} & 2021 & BC-based architecture for intelligent vaccine distribution approaches in COVID-19 pandemic situation.\\
\midrule

\cite{kabir2021application} & 2021 & Integration of Blockchain for Supply chain management. \\
\midrule

\cite{9642537} & 2021 & Blockchain-based privacy for smart healthcare management.\\
\midrule

\cite{pan2020blockchain} & 2020 &
Impact of BC on applications covering different industrial sectors.
\\
\midrule

\cite{tonnissen2020analysing} & 2020 &	
Effects of BC technology on the logistics industry and other business models. \\
\midrule

\cite{janssen2020framework} & 2020 &	
Framework for the analysis of BC integrating institutional, market, and technical factors.
\\
\midrule

\cite{xu2020effective} & 2020 & 
BC-based decentralized app for smart building system management.
\\
\midrule

\cite{9350419} & 2020 & 
Blockchained-based security for cloud storage management in IoT networks.
\\
\midrule

\cite{nawari2019blockchain} & 2019 &
Applications in the architecture, engineering, and construction industries.
\\
\midrule

\cite{wang2019blockchain} & 2019 & 
BC-enabled SCs in different application scenarios (e.g., finance, energy, etc.).
\\
\midrule

\cite{lu2019blockchain} & 2019 &
Review of the BC systems exploited in the oil and gas industry.
\\
\midrule

\cite{tijan2019blockchain} & 2019 &
Main trends of BC usage in supply chains management and logistics.
\\
\midrule

\cite{9290627} & 2019 & SDN-IoT model with NFV implementation for smart cities based on a distributed protected BC architecture.
\\
\midrule

\cite{aggarwal2019blockchain} & 2019 &  
Applications for smart communities (e.g., smart grid, transportation, healthcare).
\\
\midrule

\cite{cha2018blockchain} & 2018 & 
BC-connected gateways to maintain user privacy in IoT networks.
\\
\midrule

\cite{zhang2018ledgerguard} & 2018 & \texttt{LedgerGuard}: a tool for ledger integrity, detecting and recovering corrupted blocks.
\\ 
\midrule

\cite{griggs2018healthcare} & 2018 & Patient monitoring system using SC and private BC based on the Ethereum protocol.
\\ 
\midrule

\cite{swati2018application} & 2018 & Potential applications of BC in travel industry.
\\
\midrule

\cite{cheng2018blockchain} & 2018 & Decentralized apps and design of a certificate system based on Ethereum BC.
\\ 
\midrule

\cite{ghazali2018graduation} & 2018 & Model for academic certificate verification using BC technology.
\\ 
\midrule 

\cite{grather2018blockchain} & 2018 & \textit{BC for Education} to support protection and secure management of certificates.
\\
\midrule

\cite{dorri2017towards} & 2017 & Lightweight instance of a BC system for smart home environments.
\\
\midrule

\cite{aras2017blockchain} & 2017 & 
Current and possible future applications of BC in various domains.
\\ 
\midrule

\cite{pass2017fruitchains} & 2017 & \textit{FruitChains}: a new protocol to ensure fairness in a BC.
\\ 
\midrule

\cite{sharples2016blockchain} & 2016 &  
BC-based distributed system for educational record and reputation.
\\ 
\bottomrule
\end{tabular}
}
\end{table*}

BC technologies are used in many areas, not only in financial applications. Figure~\ref{fig:bcapp} collects some of the application fields of BC technology covering several industrial sectors.
In the following, we describe some of the most common BC applications. For the sake of completeness, other interesting ones are reported in Tab.~\ref{tab:BC-App} where recent works are summarized along with their application fields and related contributions.
Unfortunately, almost all of such systems providing smart services (and their actual applicability) are validated and evaluated in simulation environments given the high cost needed to employ them in production.

\paragraph{Bitcoin and Cryptocurrencies.}
Bitcoin is a cryptocurrency, often described as digital or virtual currency because when using it, there are no bank transactions or any physical medium. Mining is a process originated from bitcoin systems where miners verify transactions and can earn bitcoin when validating a particular number of transactions. Bitcoin is stored in wallets which are referred to as a string of letters and numbers. These can be a piece of paper, a hardware device, or online-based.
A physical bitcoin is useless if it doesn't have any private codes inside.
Starting in $2009$, the cryptocurrency unit price was only $0.01\$$; nowadays $1$ bitcoin is worth $\approx40k\$$.
Notably, the value of bitcoin has price ups and downs when exchanged with traditional currency due to politics, media hype, and country acknowledgements.
Without being exhaustive, other than bitcoin, the most important BC-based cryptocurrencies (based on their market cap) are Ethereum, Litecoin, Tether, Binance Coin, and Cardano.

\paragraph{Supply Chain Management.}
We now discuss complex supply chains that extend to all parts of the world~\cite{abeyratne2016blockchain}.
All physical products should take a journey from the seller to the buyer, and the process that allows to transfer such products is referred to as supply chain.
In this context, transparency is related to the information accessible to firms composing the provider network~\cite{francisco2018supply}.
The path to the consumer is not straightforward and sometimes there are dozens of intermediaries involved. The BC technology can potentially improve the transparency and traceability within the manufacturing supply chain through the use of the immutable record of data, distributed storage, and controlled user access~\cite{abeyratne2016blockchain}.

\begin{figure}[t]
    \centering
    \includegraphics[width=0.50\textwidth]{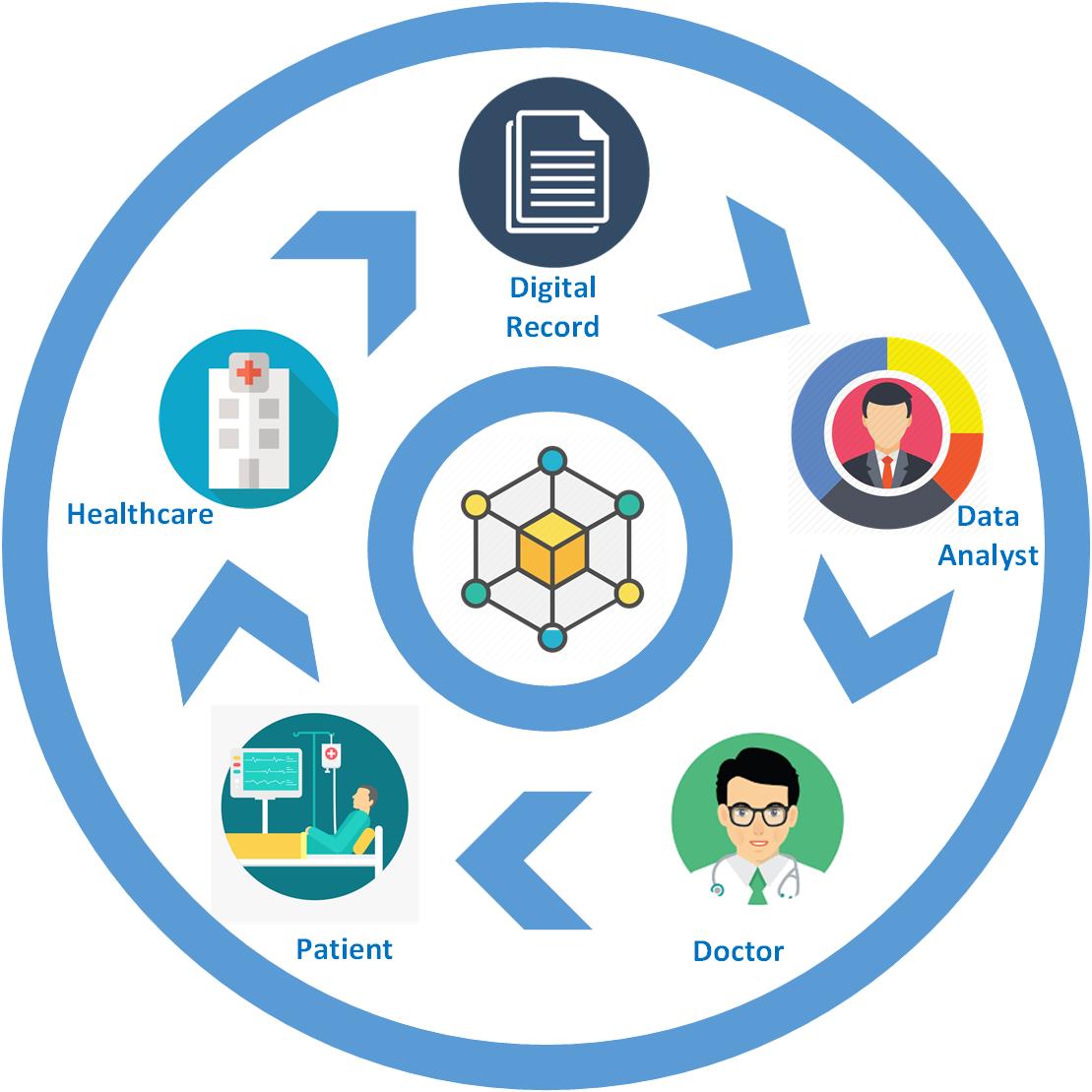}
    \caption{Exemplifying healthcare transaction using Blockchain.}
    \label{fig:Healthcare}
\end{figure}

\paragraph{Healthcare.}
Nowadays, the use of IoT in medical care is rapidly increasing because of the impressive growth of medical data.
Sharing such data is a critical aspect for improving the quality of healthcare services and reducing medical costs.
The concept of protected healthcare information has been introduced to handle secure and trusted data transactions for transmitting and storing medical data.
Indeed, though current healthcare systems bring much convenience, many obstacles still exist in practice, that hinder secure and scalable data sharing across multiple organizations, and thus limiting the development of medical decision-making and research.
Particularly, in a centralized system, there exist risks related to the single-point of attack and data leakage.
This may result in unauthorized use of patients' private data by curious organizations.
In this case, it is necessary to ensure security and privacy-protection and return the control right of data back to users to encourage data sharing.

BC-based SCs are introduced to facilitate secure analysis and management of medical sensors and to ensure data security and authorization (cf. Sec.~\ref{subsec:bc_security}) for patients and medical professionals~\cite{shi2020applications,griggs2018healthcare}.
Figure~\ref{fig:Healthcare} depicts a healthcare transaction using the BC and the different stakeholders involved, showing how BC-based systems establish a secured connection between the doctor, patient, and data analyst in healthcare organizations to manage the patient records securely. In such a way, the BC can address a variety of problems (e.g., care coordination, health data security, and interoperability issues) since---as shown in Fig.~\ref{fig:Healthcare}---it can harness the data stream to improve the quality of care by sharing medical records, protecting sensitive data from threat actors, and giving patients more control over their information. The BC can aggregate a patient's medical and prescription records from multiple sites/providers to generate a single, up-to-date aggregate record that medical professionals can comprehensively refer to when treating patients.

\paragraph{E-voting.}
E-voting is an election voting system where users cast votes through a digital system with a secure, reliable, and secret balloting over the Internet. Many countries in the world use the e-voting system for its anonymity security, reliability, integrity, and availability. The BC with the SCs emerges as a good candidate to develop a cheaper, more secure, more transparent, and easier-to-use e-voting system~\cite{yavuz2018towards}. Using BC, the main focus is to make the voting process fair and without any third-party mediation. There are many platforms to deal with the e-voting systems; one of them, Ethereum, is a widespread network for deploying such applications. The main concern in this scenario is to protect the users' identity and preserve transparency and integrity of data. To face this challenge Ethereum provides different hash values to users in the network through which it is almost impossible to identify the individuals. At the same time, each transaction is visible to everyone in the network~\cite{pareek2018voting} and can be validated: this makes it transparent to all the nodes in the network and maintains the integrity of data. Moreover, in distributed databases, data are stored in particular locations, making the data unmodifiable and thus the vote can not be manipulated.

\paragraph{Smart Contract.}
An SC is a protocol that digitally facilitates the verification, control, or execution of an agreement. It is a compromise between two or more parties in the form of a digital code. On the BC, participants run these codes which are kept in a public database and are unchangeable.
In other words, SCs are a set of rules, and the BC manages the transactions that meet them so that they can be delivered automatically without a third-party.
With a contract, some conditions needs to meet up; in this case, both rules and data are stored in the BC.
Moreover, if any illegal transaction according to the SC occurs, the BC can swiftly rollback the transaction~\cite{gatteschi2018blockchain}.
Summarizing, SC is acknowledged as a contract having as features self-execution, transparency, flexibility, and self-enforcing~\cite{wang2021ethereum}.

\begin{figure}[t]
    \centering
    \includegraphics[width=0.8\textwidth]{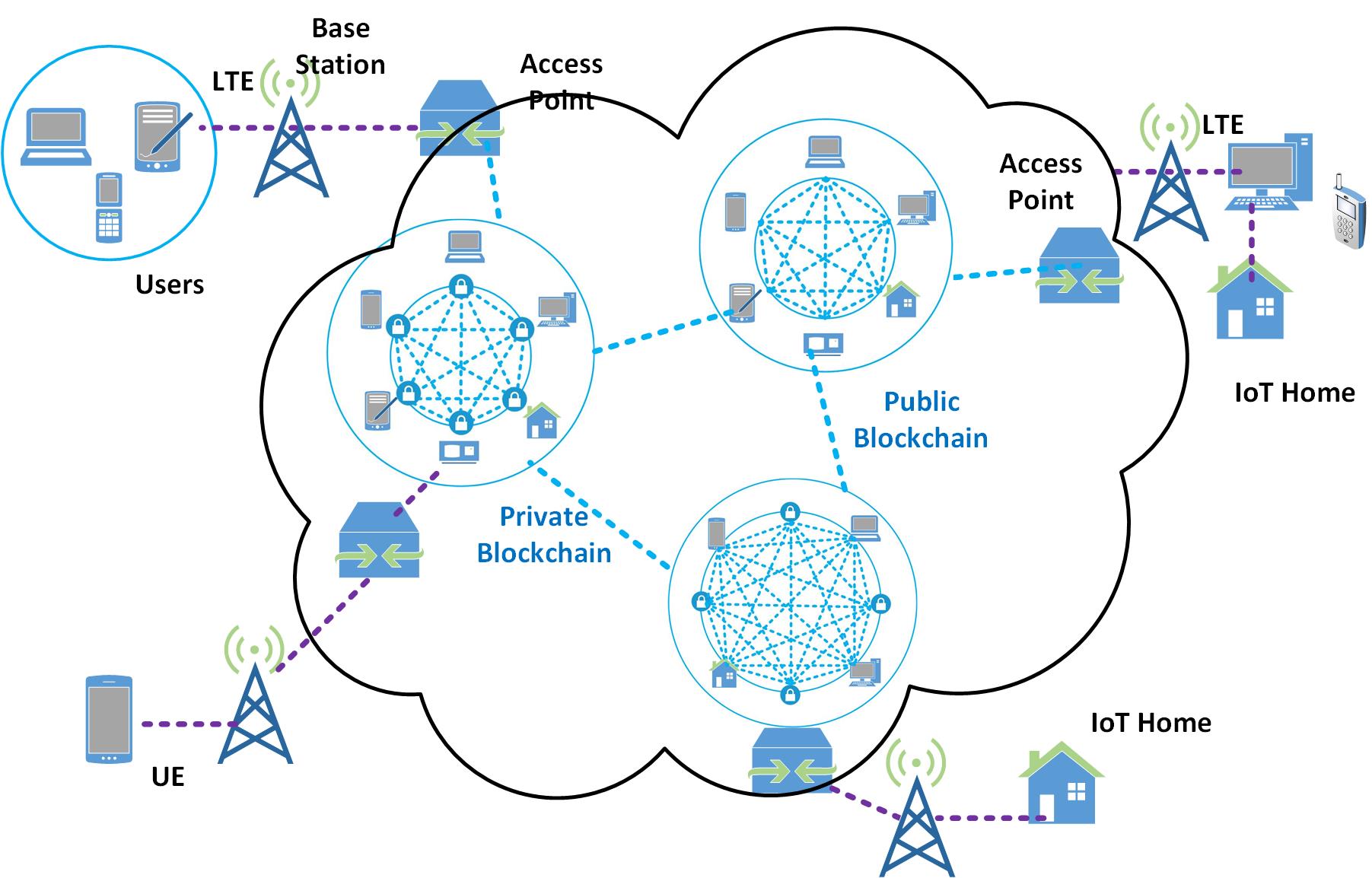}
    \caption{Integration of the Blockchain in an IoT environment.}
    \label{fig:bciots}
\end{figure}

\paragraph{Internet of Things.}
IoT presents a massive variety of devices offering possibilities for remote monitoring in numerous applications of several domains~\cite{sankaran2018towards,rahman2021study}.
IoT is widely used in smart industries, smart transportation, healthcare, military/battlefield-things, etc., and it is considered a revolution which is changing our world. Given its impressive dimensions and distributed nature, privacy and security are the main concerns~\cite{DBLP:conf/globecom/BovenziACPP20,nascita2022machine}.
Many frameworks are proposed to secure IoT environments; among them BC is increasingly used to make IoT platforms secure~\cite{Dorri2017BlockchainFI}.
In BC-based IoT systems, as shown in Fig.~\ref{fig:bciots}, private BC performs tasks efficiently and at low cost in closed environments, whereas public BC can establish connections between multiple IoT environments once at a time. In this case, management of trust is crucial and is maintained centrally without any third-party involvement.

\paragraph{Hyperledger.}
Hyperledger is an open-source collaborative
effort that has been established to advance BC technologies. It is hosted by the Linux Foundation, and it is an all-embracing co-operation platform that supports ledgers in supply chains, banking, manufacturing, IoT, finance, and technology. In simple terms, Hyperledger can be thought of as a software that enables developers all across the globe to develop BC-based solutions for particular businesses~\cite{9333523}.

\paragraph{Industry 4.0.}
Recently, industrial activities are advancing towards automation~\cite{aceto2020industry}.
Activities that do not strictly require human intervention are performed with the help of several automated technologies, and BC is one of the most important.
With its properties and the help of encrypted algorithms, nowadays it has been applied (or proposed to apply) to different industrial sectors~\cite{Rahman2020,zuo2020making}. 
For instance, in the power electronics sector, through this technology, a producer can submit an offer for a certain device, while engineers evaluate and decide whether to accept it.
In case of no matching, the producer has the possibility to resubmit the offer to reduce the mismatch with the supply~\cite{yan2017blockchain}.
Flexibility, authorization mechanisms, and protection must be guaranteed when moving to the automation process. BC allows a more sustainable and scalable ecosystem for smooth industrial operations supporting other innovative technologies such as edge computing~\cite{sitton2019new}.
Moreover, Industry 4.0 depends on data received from multiple sources which are processed to produce an outcome.
The entire framework would fail if data are not properly validated.
Also, ensuring their protection is necessary so that a third-party can not compromise the data.
In this respect, BC can provide Industry 4.0 framework with unalterable, secure, and privacy-preserving data storage~\cite{fernandez2019towards}.

\paragraph{Others Applications.}
Apart from the above-discussed applications, BC is also used in finance, business, government, asset management, insurance, personal identification, real estate, and many other fields as summarized in Fig.~\ref{fig:bcapp}.

\subsection{BC-SDN Applications}
\label{subsec:bc_sdn_applications}

The integration of BC and SDN can offer various systems some remarkable features. For example, in~\cite{okon2020blockchain}, the authors introduce a particular deployment approach for \textit{multi-operator small cells} exploiting the fusion of BC and SDN.
In detail, BC is implemented above the SDN platform that qualifies the operations within the network. These two technologies are also used in \textit{vehicular networks}, especially for security purposes because SDN provides flexibility and BC offers trust to the network.

Generally, to diminish the vulnerability of the huge number of network devices, SDN control plane is not satisfactory: this is why BC takes the place to \textit{validate insertions into flow tables}~\cite{hu2020blockchain}.
Researchers propose to send the information from network devices via BC agents which act as the mediator plane to verify insertions and are enforced between the SDN switches and SDN controller. This mechanism is responsible for vicious flows. 

The monocentric architecture of SDN exposes it to a great range of attacks; for this reason, private BC is used to manage the resources along with the other technologies~\cite{misra2020blockchain}.
In some particular contexts, SDN controllers are used to make a \textit{chain of controllers} as if each of them constitutes a block of the private BC.
Encryption is also used before storing the information and also to establish \textit{P2P connections} between SDN controllers and other devices~\cite{yazdinejad2020energy}.

A private BC is used in each domain of an SDN, while each SDN controller is associated with other controllers through a public BC. In this way, security of SDN is attained, while a significant number of works aims to improve the performance of cloud storage combined with BC-SDN.
For instance, in~\cite{sharma2017software} a \textit{distributed BC cloud architecture} based on software-defined fog nodes is presented.
In the \textit{fog layer}, the SDN controllers realize a distributed network forming a BC, while in the cloud layer, another BC is constituted with the \textit{cloud storage}.
When \textit{multiple controllers} are deployed in an SDN, BC can also harmonize their joint cooperation.

There are many other examples where these two technologies are employed in concert~\cite{qiu2018blockchain}. Indeed, although SDN has extremely valuable benefits, it requires BC support especially into the control plane to guarantee the security~\cite{shao2019blockchain}.
For example, to mitigate DDoS attacks, in~\cite{9528133,el2019cochain} is investigated the usage of \textit{SCs in the Ethereum platform}, whose rules are programmatically defined in the SCs.

\section{Challenges and Integration with Emerging Technologies}
\label{sec:FD}

In this section, we firstly discuss limitations and challenges of BC-SDN integration, then we present some emerging technologies that can be used along with the ones we have analyzed in this survey to offer valuable facilities for new applications.

\paragraph{Limitations and Challenges.}
Possible limitations of BC-SDN integration are specifically related to the real-world implementations of solutions based on it.
In fact, most of the previous works have demonstrated its functioning and benefits in a simulated environment, without evaluating its actual feasibility.
When assessing real-world implementations, the scalability of the state-of-art solutions is an open issue.
For instance, if BC-SDN is exploited for managing and securing a considerable number of nodes in an IoT scenario, it may incur severe performance degradation or even failures.
Also, the performance of an actual system could be severely impacted by the specific BC implementation leveraged since the most convenient platforms come at a huge cost in terms of implementation, processing, and energy: this is still an open challenge in both the research community and industry (e.g., finance and technology sectors).

Another open challenge of SDN is related to synchronization. More specifically, there is a lack of standardized methods to support synchronization in SDN.
Additionally, data confidentiality in BC should be carefully managed by taking into account the public visibility of BC among its users~\cite{nguyen2018survey}.
This could cause distrust in governments and organizations in widely employing such technologies when managing and sharing sensible data.

\paragraph{Blockchain for Big Data Analysis.}
\label{subsec:bc_bd}
Data science is becoming the heart of today's world.
It is a fact that individuals and corporations with huge amount of data, information, and knowledge extracted from them would have a great advantage in modern business and government.
The techniques to analyze and deal with the high dimensional data are another key topic in modern research and are commonly known as Big Data analysis. Since BC immutably stores records of information through blocks, it will provide newer services and technologies for such analyses.
Indeed, data transactions will be more secure if BC can predict some patterns about the upcoming records. Also, data would be accessible to various interested actors participating in the chain without possible errors or corruption.

\paragraph{Blockchain for Robotics.}
\label{subsec:robot_bc}
Robotics is another fundamental application field strictly related to Artificial Intelligence (AI).
AI-enabled robotics together with BC could be exploited to implement bots that execute their operations communicating with cloud/fog systems without interrupting their operability because of data corruption or manipulation (due to attacks or not).
Indeed, BC could prevent illegal instructions from being injected into the robots, thus securing both authentication and operability. The remote decision making process for robots functioning based on BC would also decrease the complexity in terms of both time and space.

\paragraph{Advanced Services via Holochain with AI-enabled BC.}
\label{subsec:holochain}
A main drawback of BC is that it is characterized by a considerable computation complexity, and it needs more memory with the increasing volume of transactions. 
Holochain is an alternative technology that could replace BC to ensure privacy and security in IoT networks, and has already gained the attention of researchers~\cite{zaman2021thinking}.
More specifically, Holochain moves from the data-centric structure typical of BC to an agent-centric one.
Indeed, Holochain blocks are independent and each possesses its ``hashchain'', without the need of reaching the consensus for data validation. Such a distributed architecture can significantly improve scalability via the proper integration of micro-services and distributed application with always-online devices (e.g., IoT sensors)~\cite{janjua2020proactive}.
If, on the one hand, Holochain can better manage the scalability problem experimented by BC due to the increase of the transactions volume, on the other hand, BC could be used in combination with Holochain to offer different services that can take advantage from both technologies. For instance, Holochain can be used to develop different distributed BC solutions that can communicate to each other.
Additionally, combining AI into such a system can improve the security control in Holochain when needed. Moreover, BC can continue to be exploited for transaction storing and authentication in smaller networks.

\paragraph{Machine and Deep Learning with Blockchain.}
\label{subsec:ml_dl_bc}
Commonly, BC adds a new block of transactions to the chain after reaching the consensus (e.g., based on PoW) and performing the authentication of every single transaction.
Authentication task can be automated using SCs that are blocks of predefined instructions allowing trusted transactions amongst the anonymous nodes without the presence of a central legal authority.
This process leveraging SCs could be replaced by Machine Learning (ML) approaches that would exploit a dynamic trained model rather than a fixed set of predefined rules.
Deep Learning (DL) is slightly different from the traditional ML paradigm.
Indeed, DL models are trained directly from input data and can attain higher accuracy especially when working with unseen data (e.g., via unsupervised or semi-supervised approaches)~\cite{aceto2019mobile}.
As seen in previous sections, IoT sensors generate huge amount of new data in real-time, while BC is successfully exploited for secure transactions.
In these systems, DL models can be deployed to perform automatic operations enhancing the performance of ML approaches and efficiently providing several automated online services.
Overall, by using DL to manage the chain, security can be also significantly enhanced~\cite{bovenzi22comparison}. Moreover, since such models work better with large amount of data, they can take advantage of the decentralized nature of BC fostering data sharing.

\section{Conclusion}
\label{sec:conclusion}

In this survey, we have investigated two different technologies which have found extensive application in network-related scenarios, namely BC and SDN.
Firstly, we have provided a thorough overview of each technology discussing also their main features, opportunities, and facilities they can provide to modern applications. Moreover, we have explored the innovative integration of these technologies---named BC-SDN---proposed to face challenges of modern networking scenarios.
Indeed, BC-SDN allows security, reliability, flexibility, cost-effective management to different applications in various networking fields.
In this regard, we have reviewed and discussed state-of-art proposals providing both motivations and details. 
Security and privacy issues are one of the most important factor driving the integration of innovative technologies, thus they have been a key focus of our discussion.
We have then analyzed such issues related to considered technologies and their integration. Actual applications of BC, SDN, and their integration are further described to provide a real-world flavor to our discussion, showing that current ``smart'' application fields can not disregard the fruitful utilization of such technologies.
Finally, the integration with further present and emerging technologies (e.g., Robotics, AI, Holochain) is also envisioned and discussed.

It is worth noticing that the integration of BC with SDN can help to shape their social impact. The improved manageability, transparency, and security of applications exploiting BC-SDN would rebuild the bridges between centralized systems and application users thanks to its tracked, audited, and (if needed) publicly accessible data.
Indeed, such peculiar features offer far-reaching possibilities for social impact, such as transaction transparency, personal data protection, legitimacy, compliance, trust, etc.
As discussed previously, the potential use cases span from financial transactions to e-voting and healthcare in which (monetary/election/medical) data can be safely stored and are instantly available to stakeholders when needed, also in case of emergency.
However, if, on the one hand, there are a lot of opportunities and advantages, on the other hand, it is hard to forecast if such a shift will happen.
Indeed, this depends on how organizations and governments will embrace such technologies: depending on its deployment, a given technology can either be disruptive or transformative.

In view of the outcomes of the present survey, we think that the fruitful trend of jointly using BC and SDN will continue and produce positive synergy.
Several domains and applications would benefit from such an integration, and both researchers and practitioners working in such domains should continue exploring BC-SDN integration with the aim of providing advantages such as management flexibility, scalability and data flow verification.
In fact, there is room for further research effort for the improvement of security and performance of BC-SDN also in view of upcoming use cases and (sadly) possible threats.

Future avenues that should be investigated are the technical challenges underlining the considered technologies when applied in scenarios having particular constraints in terms of scalability and computational efficiency.




%
%
\bibliographystyle{ieeetr}
\bibliography{sample}

\clearpage

\section*{Authors' Biography}

\textbf{Anichur Rahman} received the B.Sc. and M.Sc degree in Computer Science and Engineering from Mawlana Bhashani Science and Technology University, Bangladesh in 2017 and 2020 respectively. Currently, he is working as a Lecturer at Computer Science and Engineering  National Institute of Textile Engineering and Research, Bangladesh. His research interests include the Internet of Things, Blockchain, Software Defined Networking, Machine Learning, 5G, Industry 4.0, and Data Science.
\vspace{10pt}

\noindent\textbf{Antonio Montieri} is an Assistant Professor at DIETI of the University of Napoli Federico II. He has received his Ph.D. degree in Information Technology and Electrical Engineering in April 2020 from the same University. His work concerns network measurements, (encrypted and mobile) traffic classification, traffic modeling and prediction, and monitoring of cloud network performance. Antonio has co-authored 35 papers in international journals and conference proceedings.
\vspace{10pt}

\noindent\textbf{Dipanjali Kundu} received the B.Sc. degree in Computer Science and Engineering from Chittagong University of Engineering and Technology, Bangladesh in 2018. Currently, she is working as a Lecturer at Computer Science and Engineering, National Institute of Textile Engineering and Research, Bangladesh. Her research interests include Machine Learning, Human Computer Interaction, Internet of Things, Blockchain, Software Defined Networking, 5G, Industry 4.0, and Robotics.
\vspace{10pt}

\noindent\textbf{Md. Razaul Karim} received the B.Sc. degree in Computer Science and Engineering from Mawlana Bhashani Science and Technology University, Tangail, Bangladesh in 2020. The main interests of his research are Machine Learning, Computer Vision, and Image Processing. He is also keen on Blockchain.
\vspace{10pt}

\noindent\textbf{Md. Jahidul Islam} received the B.Sc. and M.Sc. degrees in Computer Science and Engineering from Jagannath University, Dhaka, in 2015 and 2017 respectively. Currently, he is working as a Lecturer and Program Coordinator at Computer Science and Engineering, Green University of Bangladesh. His research interests include Internet of Things, Blockchain, Network Function Virtualization, Software Defined Networking, 5G, Industry 4.0, Machine Learning, and Wireless Mesh Networking.
\vspace{10pt}

\noindent\textbf{Umme Sara} received her B.Sc. in 2012 and M.Sc. in 2014 both from the Jahangirnager University under the discipline of Computer Science and Engineering. Currently, she is working as an Assistant Professor and Head at the Department of Computer Science and Engineering, National Institute of Textile Engineering and Research, Bangladesh. Her current research interests include Machine Learning, Digital Image Processing, Computer Vision Systems, Data Science, IoT, and Blockchain.
\vspace{10pt}

\noindent\textbf{Alfredo Nascita} is a PhD Student in Information Technology and Electrical Engineering at DIETI, University of Napoli Federico II. He received his M.S. Laurea Degree in Computer Engineering from the same University in March 2021. His research interests include traffic classification, machine and deep learning, and explainable artificial intelligence.
\vspace{10pt}

\noindent\textbf{Antonio Pescapé} is a Full Professor of computer engineering at the University of Napoli Federico II. His work focuses on measurement, monitoring, and analysis of the Internet. He has co-authored more than 200 conference and journal papers, he is the recipient of a number of research awards. Also, he has served as an independent reviewer/evaluator of research projects/project proposals co-funded by a number of governments and agencies.
\vspace{10pt}

\end{document}